\documentclass[twocolumn]{aastex631}
\usepackage{xspace}

\newcommand{\Lsun}{L_\odot}
\newcommand{\Msun}{M_\odot}
\newcommand{\MJyr}{\MJ\,\mathrm{yr}^{-1}}

\newcommand{\flux}{W\,m$^{-2}$~$\mu$m$^{-1}$}

\newcommand{\MJ}{\ensuremath{M_\mathrm{J}}\xspace}
\newcommand{\RJ}{\ensuremath{R_\mathrm{J}}\xspace}

\newcommand{\Mdot}{\ensuremath{\dot{M}}\xspace}

\newcommand{\Mp}{\ensuremath{M_\mathrm{p}}\xspace}
\newcommand{\Rp}{\ensuremath{R_\mathrm{p}}\xspace}

\newcommand{\Teff}{\ensuremath{T_\mathrm{eff}}\xspace}
\newcommand{\logg}{\ensuremath{\log{g}}\xspace}

\newcommand{\Tdisk}{\ensuremath{T_\mathrm{disk}}\xspace}
\newcommand{\Rdisk}{\ensuremath{R_\mathrm{disk}}\xspace}

\newcommand{\ff}{\ensuremath{f_\mathrm{fill}}\xspace}

\newcommand{\wavel}{4--5~$\mu$m\xspace}
\newcommand{\gq}{GQ~Lup~B\xspace}

\newcommand{\ffarcs}{\mbox{\ensuremath{.\!\!^{\prime\prime}}}}

\received{August 09, 2021}
\revised{September 20, 2021}
\accepted{October 01, 2021}
\submitjournal{AJ}

\shorttitle{Characterizing the protolunar disk of the accreting companion GQ~Lupi~B}
\shortauthors{Stolker et al.}

\begin{document}

\title{Characterizing the protolunar disk of the accreting companion GQ~Lupi~B\footnote{Based on observations collected under ESO programmes \mbox{0101.C-0502(B)}, \mbox{0102.C-0649(A)}, and \mbox{0103.C-0524(A)}.}}

\author[0000-0002-5823-3072]{Tomas Stolker}
\affiliation{Leiden Observatory, Leiden University, Niels Bohrweg 2, 2333 CA Leiden, The Netherlands}

\author[0000-0001-5130-9153]{Sebastiaan Y. Haffert}
\altaffiliation{NASA Hubble Fellow}
\affiliation{Steward Observatory, University of Arizona, 933 North Cherry Avenue, Tucson, AZ 85721, USA}

\author[0000-0002-3239-5989]{Aurora Y. Kesseli}
\affiliation{Leiden Observatory, Leiden University, Niels Bohrweg 2, 2333 CA Leiden, The Netherlands}

\author[0000-0003-1520-8405]{Rob G. van Holstein}
\affiliation{Leiden Observatory, Leiden University, Niels Bohrweg 2, 2333 CA Leiden, The Netherlands}
\affiliation{European Southern Observatory, Alonso de C\'{o}rdova 3107, Casilla 19001, Vitacura, Santiago, Chile}

\author[0000-0003-0568-9225]{Yuhiko Aoyama}
\affiliation{Department of Earth and Planetary Science, University of Tokyo, 7-3-1 Hongo, Bunkyo-ku, Tokyo 113-0033, Japan}
\affiliation{Institute for Advanced Study, Tsinghua University, Beijing 100084, People’s Republic of China}
\affiliation{Department of Astronomy, Tsinghua University, Beijing 100084, People’s Republic of China}

\author[0000-0003-4359-8797]{Jarle Brinchmann}
\affiliation{Leiden Observatory, Leiden University, Niels Bohrweg 2, 2333 CA Leiden, The Netherlands}
\affiliation{Instituto de Astrof\'{i}sica e Ci\^{e}ncias do Espa\c{c}o, Universidade do Porto, CAUP, Rua das Estrelas, PT4150-762 Porto, Portugal}

\author[0000-0001-7255-3251]{Gabriele Cugno}
\affiliation{Institute for Particle Physics and Astrophysics, ETH Zurich, Wolfgang-Pauli-Strasse 27, 8093 Zurich, Switzerland}

\author[0000-0001-8627-0404]{Julien H. Girard}
\affiliation{Space Telescope Science Institute, 3700 San Martin Drive, Baltimore MD, 21218, USA}

\author[0000-0002-2919-7500]{Gabriel-Dominique Marleau}
\affiliation{Institut f\"ur Astronomie und Astrophysik, Universit\"at T\"{u}bingen, Auf der Morgenstelle 10, 72076 T\"{u}bingen, Germany}
\affiliation{Physikalisches Institut, Universit\"{a}t Bern, Gesellschaftsstrasse 6, 3012 Bern, Switzerland}
\affiliation{Max-Planck-Institut f\"ur Astronomie, K\"{o}nigstuhl 17, 69117 Heidelberg, Germany}

\author[0000-0003-1227-3084]{Michael R. Meyer}
\affiliation{Department of Astronomy, University of Michigan, 1085 South University Avenue, Ann Arbor, MI 48109-1107, USA}

\author[0000-0001-9325-2511]{Julien Milli}
\affiliation{Universit\'{e} Grenoble Alpes, IPAG, F-38000 Grenoble, France}

\author[0000-0003-3829-7412]{Sascha P. Quanz}
\affiliation{Institute for Particle Physics and Astrophysics, ETH Zurich, Wolfgang-Pauli-Strasse 27, 8093 Zurich, Switzerland}

\author[0000-0003-1624-3667]{Ignas A.\,G. Snellen}
\affiliation{Leiden Observatory, Leiden University, Niels Bohrweg 2, 2333 CA Leiden, The Netherlands}

\author[0000-0002-9276-8118]{Kamen O. Todorov}
\affiliation{Anton Pannekoek Institute for Astronomy, University of Amsterdam, Science Park 904, 1090 GE Amsterdam, The Netherlands}

\correspondingauthor{Tomas Stolker}
\email{stolker@strw.leidenuniv.nl}

\begin{abstract}

\gq is a young \added{and accreting}, substellar companion that appears to drive a spiral arm in the circumstellar disk of its host star. \deleted{Hydrogen emission lines had been identified with optical photometry and near-infrared spectroscopy, indicating that the object is still forming through the accretion of gas.}\explain{Removed to shorten the abstract to less than 250 words} We report high-contrast imaging observations of \gq with VLT/NACO at \wavel and medium-resolution integral field spectroscopy with VLT/MUSE. The optical spectrum is consistent with an M9 spectral type, shows characteristics of a low-gravity atmosphere, and exhibits strong H$\alpha$ emission. The $H$~--~$M'$ color is $\gtrsim$1~mag redder than field dwarfs with similar spectral types and a detailed analysis of the spectral energy distribution (SED) from optical to mid-infrared wavelengths reveals excess emission in the $L'$, NB4.05, and $M'$ bands. The excess flux is well described by a blackbody component with $\Tdisk \approx 460$~K and $\Rdisk \approx 65~\RJ$ and is expected to trace continuum emission from small grains in a protolunar disk. We derive an extinction of $A_V \approx 2.3$~mag from the broadband SED with a suspected origin in the vicinity of the companion. We also combine 15~yr of astrometric measurements and constrain the mutual inclination with the circumstellar disk to $84 \pm 9$~deg, indicating a tumultuous dynamical evolution or a stellar-like formation pathway. From the measured H$\alpha$ flux and the estimated companion mass, $\Mp \approx 30~\MJ$, we derive an accretion rate of $\Mdot \approx 10^{-6.5}~\MJyr$. We speculate that the disk is in a transitional stage in which the assembly of satellites from a pebble reservoir has opened a central cavity while \gq is in the final stages of its formation.

\end{abstract}


\section{Introduction}
\label{sec:introduction}

Directly imaged planets and brown dwarfs are an intriguing population of low-mass objects that have been discovered at large distance from their star \citep[e.g.,][]{oppenheimer2009,bowler2016}. Their super-Jupiter masses and wide orbits challenge planet formation theories due to the long timescales associated with a bottom-up formation of rocky cores and subsequent gas accretion in circumstellar disks (CSDs). Some of these objects may instead have formed through a stellar-like formation pathway---in particular those with a high companion-to-star mass ratio and/or on wide orbits.

Large-scale surveys have revealed that planets are typically found at smaller separations around intermediate-mass stars whereas brown dwarfs are detected on larger orbits around lower-mass stars \citep[e.g.,][]{nielsen2019,vigan2021}. This may suggest a stellar formation scenario for substellar-mass objects on wide orbits, which is in line with eccentricity constraints inferred from their orbital dynamics \citep{bowler2020}. The chemical composition of their atmospheres provides another glimpse into their formation history \citep[e.g.,][]{oeberg2011,madhusudhan2014}. Atmospheric retrievals are starting to reveal signatures of the physical and chemical processes that are at play during formation, for example through the spectral inference of the C/O ratio and metallicity, which could point to non-solar chemical compositions \citep[e.g.,][]{lee2013,molliere2020}.

The youngest (i.e., $\lesssim$10~Myr) of the directly imaged, planetary-mass companions (PMCs) provide a direct window to the formation of giant planets and brown dwarfs and their circumplanetary/substellar disk characteristics. The PDS~70 planets have become the signpost for empirical studies on embedded protoplanets \citep[e.g.,][]{keppler2018,mueller2018}. These Jupiter-mass planets have opened a gap at 20--30~au in the CSD while accreting gas and dust from their environment \citep[e.g.,][]{haffert2019,wang2020,stolker2020b}. Other accreting PMCs \citep[see Table~2 in][for an overview]{wu2017b} typically orbit further away from their star ($\gtrsim$100~au) and the connection with the CSD is less clear \citep[e.g.,][]{ireland2011,kraus2014}. The spectral energy distributions (SEDs) of some of these objects are unusually red which can be evidence for a dusty disk \citep[e.g.,][]{bailey2013,wu2015}---in line with expectations from isolated, planetary and substellar objects---that is expected to serve as the formation site of satellites. The study of the accretion and disk characteristics of both populations (i.e., embedded in versus detached from the CSD) may reveal clues about possible differences in their formation pathways and the processes by which giant planets and brown dwarfs accumulate a gaseous envelope.

One such directly-imaged substellar object is \gq, which was discovered by \citet{neuhauser2005} orbiting a K7Ve-type T~Tauri star \citep{herbig1977} with a well studied CSD \citep[e.g.,][]{dai2010,mcclure2012}. The mass of \gq has been debated ever since, with inferred masses of $\sim$10--$40~\MJ$ \citep[e.g.,][]{marois2007,mcelwain2007,seifahrt2007}. Its near-infrared (NIR) spectrum has been classified as mid~M to early~L \citep{mcelwain2007} with $\Teff \approx 2500$~K and $\logg \approx 4$ (see overview in Table~1 by \citealt{lavigne2009}). \gq was also detected with optical photometry. This revealed H$\alpha$ emission that has been linked with accretion \citep{marois2007,zhou2014,wu2017a}, in line with the detection of Pa$\beta$ emission by \citet{seifahrt2007}. In addition to the atmosphere, the orbit has been analyzed by several authors \citep{janson2006,neuhauser2008,ginski2014} with the most recent constraints pointing to an inclination of $\sim$60~deg and semi-major axis of 100-150~au \citep{schwarz2016}. The companion has a low projected spin velocity which could suggest that it is still gaining angular momentum \citep{schwarz2016} at the age of 2--5~Myr \citep{donati2012}. Recently, a second companion, GQ~Lup~C, was detected at a projected separation of $\sim$2400~au \citep{alcala2020}. This low-mass star is also surrounded by a dusty accretion disk, as inferred from UV and IR excess, with a somewhat similar orientation to the disk of GQ~Lup \citep{lazzoni2020}. \added{Apart from GQ~Lup~B and~C, who are orbiting at a large separation, there is also a gap detected in the CSD at $\sim$10~au which could be evidence for a hidden planet on a solar system scale \citep{long2020}.}

In this work, we report the first detection of \gq \replaced{at \wavel}{with \wavel imaging} and with optical spectroscopy. Mid-infrared (MIR) wavelengths are sensitive to effects of clouds and carbon chemistry but are also a powerful probe for the thermal emission from circumplanetary/substellar material. With the optical spectrum, we search for emission lines that carry accretion signatures. We analyze the \wavel photometry and optical spectroscopy in combination with archival, medium-resolution $JHK$-band spectra to detect and characterize excess emission from the disk around \gq and constrain the atmospheric and extinction properties.

\section{Observations}
\label{sec:observations}

\subsection{Mid-infrared Imaging with VLT/NACO}
\label{sec:naco_imaging}

We observed GQ~Lup with the Very Large Telescope (VLT) on Cerro Paranal in Chile as part of the MIRACLES survey (ESO program ID: 0102.C-0649(A)). The program aims at the systematic characterization of directly imaged planets and brown dwarfs at 4--5~$\mu$m \citep{stolker2020a} by using the adaptive optics-assisted imaging capabilities of NACO \citep{lenzen2003,rousset2003}. The observations were carried out with the NB4.05 (Br$\alpha$; $\lambda_0 = 4.05$~$\mu$m, $\Delta\lambda = 0.06$~$\mu$m\footnote{The FWHM of the NB4.05 filter has been incorrectly listed on the ESO website as 0.02~$\mu$m so we revised the value to 0.06~$\mu$m.}) and $M'$ ($\lambda_0 = 4.78$~$\mu$m, $\Delta\lambda = 0.59$~$\mu$m) filters on the nights of 2019~Mar~14 and 2019~Mar~08, respectively. A description of the observing strategy can be found in \citet{stolker2020a}, but we will provide a few specific details here.

With the NB4.05 filter, we used a detector integration time (DIT) of 1~s and obtained a total of 1400 frames. In addition, we obtained 600 frames with a shorter DIT of 0.2~s for calibration purposes. The parallactic rotation was only 11.6~deg but sufficient for a detection of \gq. The atmospheric conditions during the observations were stable with a seeing of $0\ffarcs78 \pm 0\ffarcs04$ and a coherence time of $5.0 \pm 0.8$~ms. The angular resolution, as measured from the PSF, was 116~mas (1~FWHM) and the stellar flux in the unsaturated exposures varied by 2.8\%.

Similarly, we observed GQ~Lup with the $M'$ filter but with a DIT of 50~ms to prevent saturation by the high thermal background emission. We obtained a total of 34400 frames with a comparable seeing, $0\ffarcs85 \pm 0\ffarcs04$, and coherence time, $4.3 \pm 0.3$~ms as the NB4.05 observations. The total parallactic rotation was 15.6~deg and the angular resolution was 137~mas. The stellar flux varied by 3.9\% across the full stack of frames.

\subsection{Integral Field Spectroscopy with VLT/MUSE}
\label{sec:muse_observation}

GQ~Lup was also observed with the MUSE integral field spectrograph (IFS) on the night of 2019~Apr~19 (ESO program ID: 0103.C-0524(A)). MUSE operates in the visible (4800--9300~\AA) with a resolving power of $R = 1740-3450$ \citep{bacon2010}. The instrument is mounted on Unit Telescope~4 (UT4) of the VLT and therefore benefits from the adaptive optics facility \citep[AOF;][]{arsenault2008,strobele2012}. We used the narrow field mode (NFM) which leverages the laser tomography adaptive optics system (GALACSI) of the AOF and samples the field with (25~mas)$^2$~spaxel$^{-1}$ to enable high-angular resolution observations.

The observations were obtained with field tracking and comprised 12 exposures of GQ~Lup with a DIT of 139~s. For each subsequent exposure, the derotator offset was increased by 90~deg to reduce detector artifacts and improve the sampling of the line spread function (LSF). The atmospheric conditions were excellent with a seeing of $0\ffarcs32 \pm 0\ffarcs04$ and a coherence time of $16 \pm 2$~ms but the target was observed at a somewhat high airmass of 1.18--1.40. Earlier in the night, a single exposure with a DIT of 240~s was obtained of the standard star GD~108, which is a B-type subdwarf \citep{greenstein1969}. The conditions were a bit poorer but still good with a seeing of 0\ffarcs64 and a coherence time of 9~ms. The calibration target was however observed at a smaller airmass of 1.07.

\section{Data Reduction}
\label{sec:data_reduction}

\subsection{NACO}
\label{sec:naco_data_reduction}

\subsubsection{Image Processing}
\label{sec:naco_processing}

The processing and calibration of the NACO datasets were done with version~0.8.3 of \texttt{PynPoint}\footnote{\url{https://pynpoint.readthedocs.io}} \citep{stolker2019}, which applies an implementation of full-frame principal component analysis \citep[PCA;][]{amara2012,soummer2012} to remove the stellar point spread function (PSF). We followed mostly the procedure as outlined in \citet{stolker2020a}, but provide a few details here.

While a background subtraction based on the mean of the adjacent data cubes was sufficient to suppress the bright background flux, we applied an additional PCA-based background subtraction to remove striped detector artifacts in several of the frames. We followed the approach by \citet{hunziker2018} and subtracted 3 (NB4.05) and 2 ($M'$) principal components (PCs) of the background emission from the data. This lowered sufficiently the quasi-static detector noise on visual inspection. Before calibrating the data, we collapsed subsets of 2 (NB4.05) and 67 ($M'$) of the pre-processed frames to end up with a stack of 687 (NB4.05) and 499 ($M'$) frames.

\subsubsection{Relative Calibration}
\label{sec:naco_calibration}

\begin{deluxetable*}{lccccc}
\tablecaption{Photometry of \gq}
\label{table:photometry}
\tablehead{
\colhead{Filter} & \colhead{Contrast} & \colhead{GQ Lup} & \colhead{Apparent magnitude} & \colhead{Absolute magnitude} & \colhead{Flux} \\
& \colhead{(mag)} & \colhead{(mag)} & \colhead{(mag)} & \colhead{(mag)} & \colhead{(W m$^{-2}$ $\mu$m$^{-1}$)}
}
\startdata
NACO NB4.05 & $6.54 \pm 0.05$ & $5.75 \pm 0.04$ & $12.29 \pm 0.06$ & $6.36 \pm 0.06$ & $(4.80 \pm 0.28) \times 10^{-16}$ \\
NACO $M'$ & $6.50 \pm 0.05$ & $5.47 \pm 0.06$ & $11.97 \pm 0.08$ & $6.03 \pm 0.08$ & $(3.52 \pm 0.25) \times 10^{-16}$ \\
\enddata
\end{deluxetable*}

The flux and position of \gq were calibrated relative to GQ~Lup by injecting negative copies of the unsaturated PSF to minimize the flux at the companion position \citep[see][for details]{stolker2020a}. We first retrieved the parameters as function of PCs to inspect the dependence. Next, we fixed the number of subtracted PCs to 6 (NB4.05) and 10 ($M'$) and use the Markov chain Monte Carlo (MCMC) ensemble sampler \texttt{emcee} \citep{foreman2013} to determine the best-fit values and statistical errors (assuming Gaussian noise). Finally, we estimated the potential bias and systematic errors (e.g., due to speckle and background residuals) by injecting and retrieving artificial sources.

The photometric and astrometric results are listed in Tables~\ref{table:photometry} and \ref{table:astrometry}. The separations and positions angles are consistent with each other within the 1--2$\sigma$ errors. The photometric error budget of NB4.05 also includes a calibration error of 0.03~mag which is calculated from the unsaturated NB4.05 exposures while the stellar $M'$ flux remained unsaturated in all frames.

\begin{deluxetable}{lcccc}
\tablecaption{Astrometry of \gq}
\label{table:astrometry}
\tablehead{
\colhead{MJD} & \colhead{Instrument} & \colhead{Filter} & \colhead{Separation} & \colhead{Position angle} \\
\nocolhead{Number} & \nocolhead{Number} & \nocolhead{Name} & \colhead{(mas)} & \colhead{(deg)}
}
\startdata
58345.1 & SPHERE & B\_H & $711.6 \pm 2.4$ & $278.27 \pm 0.24$ \\
58551.3 & NACO & $M'$ & $720.3 \pm 3.4$ & $277.90 \pm 0.18$ \\
58557.2 & NACO & NB4.05 & $715.0 \pm 2.8$ & $277.83 \pm 0.17$ \\
58593.1 & MUSE & -- & $701 \pm 20$ & $278.3 \pm 1.2$ \\
\enddata
\tablecomments{The true north, $-1.75 \pm 0.08$~deg, and pupil offset, $-135.99 \pm 0.11$~deg, for SPHERE/IRDIS have been adopted from \citet{maire2016b} and the true north for NACO, $-0.44 \pm 0.10$~deg, has been adopted from \citet{cheetham2019}. The listed position angles have been corrected for these offsets and the uncertainties have been propagated into the error budget.}
\end{deluxetable}

\subsubsection{Absolute Flux Calibration}
\label{sec:naco_flux}

GQ~Lup has a CSD so its brightness at \wavel is affected by IR excess. Therefore, to convert the contrast values of \gq into apparent magnitudes, we calculated the flux of GQ~Lup in the NB4.05 and $M'$ filters by simply fitting the 2MASS $H$ and $K_\mathrm{s}$ and WISE $W1$ and $W2$ fluxes with a power-law function in log-log space, that is, $\log f_\lambda = a + b x^c$, where $x = \log \lambda$ and with $f_\lambda$ in W~m$^{-2}$~$\mu$m$^{-1}$ and $\lambda$ in $\mu$m. We used the Bayesian framework (see Sect.~\ref{sec:modeling} for details) in the \texttt{species}\footnote{\url{https://species.readthedocs.io}} toolkit \citep{stolker2020a} to determine the best-fit parameters and uncertainties: $a = -11.80 \pm 0.07$, $b = -1.81 \pm 0.06$, and $c = 1.38 \pm 0.17$.

The posterior distributions were then propagated into synthetic magnitudes by folding the power-law spectra with the filter profiles. From this, we derived $5.75 \pm 0.04$~mag in NB4.05 and $5.47 \pm 0.06$~mag in $M'$ for GQ~Lup, which then gave the apparent magnitudes of the companion. The absolute magnitudes were calculated by adopting the \emph{Gaia} distance of $154.1 \pm 0.7$~pc \citep{gaia2016,gaia2021}. Finally, the magnitudes were converted into fluxes with \texttt{species} by using a flux-calibrated spectrum of Vega \citep{bohlin2007} and setting its magnitude to 0.03~mag for all filter. The calibrated photometry is provided in Table~\ref{table:photometry}.

\begin{figure*}
\centering
\includegraphics[width=\linewidth]{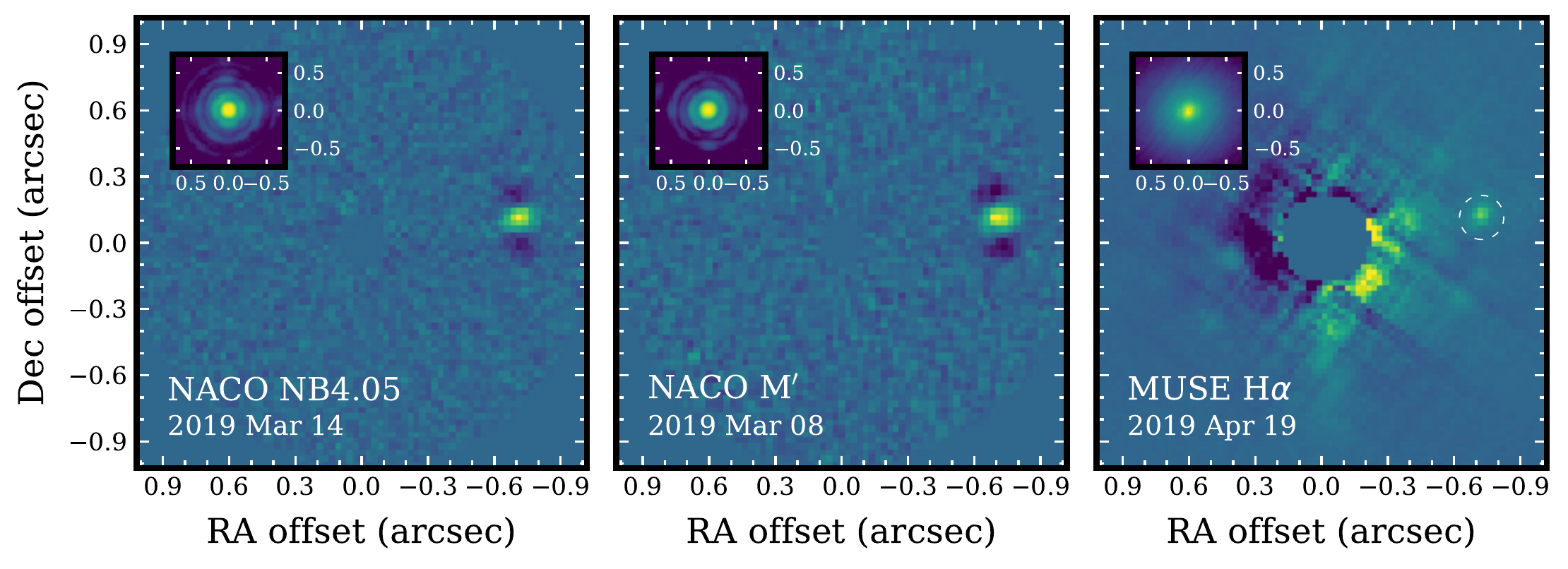}
\caption{Detection of \gq with the NACO NB4.05 (\emph{left panel}) and $M'$ (\emph{central panel}) filters, and with MUSE at H$\alpha$ (\emph{right panel}). The images show the mean-combined residuals of the PSF subtraction on a linear scale. The insets display the unsaturated PSF fluxes of GQ~Lup on a logarithmic scale. The color scales have been normalized for best visibility. North is up and east is left in all images.}
\label{fig:images}
\end{figure*}

\subsection{MUSE}
\label{sec:muse_data_reduction}

The reduction and calibration of the MUSE data was done in a similar way as outlined by \citet{haffert2020}. We used version 2.8.2 of the MUSE pipeline recipes for EsoRex \citep{weilbacher2020}. The pipeline performs a background subtraction, flat field correction, 2D spectral extraction, wavelength calibration (by using the internal arc lamps), and a telluric correction. The PSF of MUSE is not Nyquist sampled so we centered the images only with pixel precision (i.e., 25~mas) to avoid inaccuracies due to interpolation. The final product is a cleaned and flux-calibrated (by using the spectrum of the standard star) data cube, which was used for the spectral extraction of \gq.

Some spurious features were visible in the spectrum due to an imperfect correction of the telluric transmission, which may have been caused by the difference in airmass ($\sim$0.1--0.3) with the standard star. The same residuals were also seen in the spectrum of GQ~Lup so we applied a second correction by normalizing the MUSE spectrum with a PHOENIX model spectrum \citep{husser2013}, for which we adopted $\Teff = 4300$~K and $\logg = 3.7$ from \citet{donati2012}, while masking and interpolating regions with emission lines. The derived scaling correction was then applied to the spectrum of \gq which effectively removed the remaining telluric residuals.

After image registration and calibration, we removed the stellar halo by subtracting an azimuthally-averaged radial profile in steps of 12.5~mas. Additionally, a second order 2D polynomial was fitted around \gq to subtract remaining, low-frequency structures in the images. The spectrum of \gq was then extracted by fitting a shifted and normalized copy of the stellar PSF at the position of the companion. For each wavelength, the flux was estimated as the weighted (based on the PSF template) sum of the pixels in an aperture with a diameter of 150~mas, therefore also accounting for the \deleted{low-SNR} flux contributions outside the aperture \added{that have a low signal-to-noise ratio (SNR)}. The noise was estimated from the scatter of the extracted fluxes between 6300 and 7000~\AA\ by sampling random errors. In that regime, the continuum of \gq lies below the noise floor and there are limited residuals of GQ~Lup.

\subsection{Archival VLT/SPHERE $H$-band Imaging}
\label{sec:archival_sphere}

We also reanalyzed the archival SPHERE \citep{beuzit2019} $H$-band imaging dataset (ESO program ID: 0101.C-0502(B)) that was published by \citet{vanholstein2021}, in order to extract the position of the companion from this 2018 epoch (see orbit fit in Sect.~\ref{sec:orbit}). The data were obtained with the dual-beam polarimetric imaging mode \citep{deboer2020,vanholstein2020} of the IRDIS camera \citep{dohlen2008} but we only used the Stokes~I images for the astrometry. These polarimetric observations were done in pupil-tracking mode \citep[see][]{vanholstein2017}, resulting in a field rotation of only $6\deg$ but the companion is sufficiently bright to be robustly detected at a separation of $\approx$0\ffarcs7.

We reduced the data with \texttt{IRDAP}\footnote{\url{https://irdap.readthedocs.io}} \citep{vanholstein2017,vanholstein2020} and measured the relative position with \texttt{PynPoint} \citep{stolker2019}, similar to \citet{vanholstein2021} and the calibration in Sect.~\ref{sec:naco_calibration}. Given the limited field rotation, we subtracted the PSF with 2~PCs and used artificial sources to obtain the best-fit position and error bars. The centering of the frames was achieved with the satellite spots in the dedicated calibration frames to locate the star behind the coronagraph. There was a $\sim$0.2~pixel difference in the position of the star between the frames obtained at the start and end, which we adopted in the astrometric error budget as the centering precision.

The results from the astrometry are listed in Table~\ref{table:astrometry}. The errors are dominated by the accuracy of the stellar position and the true north correction. Additionally, we cross-checked the astrometric extraction by fitting a 2D Gaussian directly to the companion flux in the pre-processed frames. In that case, we obtained a very similar result with differences of 0.4~mas in separation and 0.01~deg in position angle, which is small ($\sim$10\%) compared to the total errors.

\section{Results}
\label{sec:results}

\subsection{Mid-infrared and Optical Detection of GQ~Lup~B}
\label{sec:detection}

The mean-combined residuals from the PSF subtraction with PCA of the NACO datasets are shown in the left and central panels of Fig.~\ref{fig:images}. The PSF model was created by projecting each frame of the pre-processed data onto the first 3 PCs (for both filters). \gq is relatively bright (i.e., $\sim$12~mag) at \wavel so the object is clearly detected west of the central star both in NB4.05 and $M'$. At a separation of 715--720~mas and a position angle of $\sim$278~deg (see Table~\ref{table:astrometry}), the separation has decreased by $\sim$5~mas and the position angle increased by $\sim$1~deg with respect to the last epoch by \citet{ginski2014} from 2012. The orbital architecture of \gq will be analyzed in more detail in Sect.~\ref{sec:orbit}.

The right panel of Fig.~\ref{fig:images} shows the MUSE image at H$\alpha$ after PSF subtraction. Some high-frequency stellar residuals remained present but these noise features impact mainly the blue end ($\lambda < 5500$~\AA) of the spectrum while beyond 6300~\AA\ the residuals are relatively small at the separation of \gq. The extracted spectrum will be presented and analyzed in Sect.~\ref{sec:muse_analysis}.

\subsection{Color and Magnitude Comparison}
\label{sec:color_magnitude}

\begin{figure*}
\centering
\includegraphics[width=0.75\linewidth]{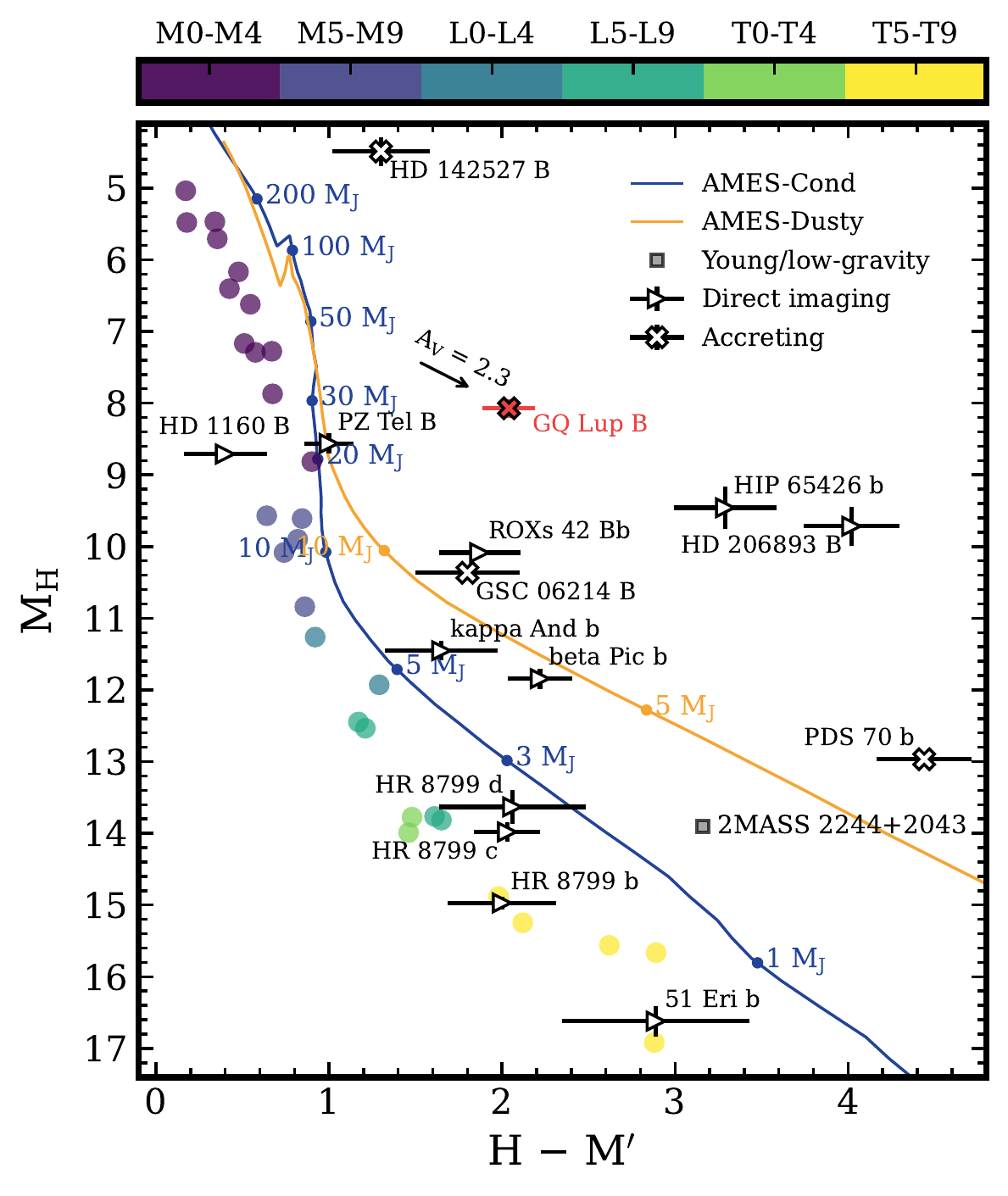}
\caption{Color--magnitude diagram of M$_H$ versus $H$~--~$M'$. The regular field dwarfs are color-coded by M, L, and T spectral types (see discrete colorbar), the young and low-gravity objects are indicated with a \emph{gray square}, and the directly imaged objects are labeled individually. \gq is highlighted with a \emph{red cross}. The \emph{blue} and \emph{orange} lines show the synthetic colors computed from isochrones at \replaced{5~Myr}{3~Myr}. \replaced{The \emph{black arrows} in the bottom left corner are reddening vectors for MgSiO$_3$ grains that were calculated for a log-normal size distribution with mean radii of 0.1 and 1~$\mu$m, a geometric standard deviation of 2, and $A_H$ of 1 and 2~mag, respectively.}{The \emph{black arrow} is a reddening vector for the extinction, $A_V = 2.3$, that is inferred from the SED of \gq (see Sect.~\ref{sec:model_atmos} for details).}}
\label{fig:color_magnitude}
\end{figure*}

Color--magnitude diagrams (CMDs) provide an insightful approach for comparing the photometric characteristics of substellar and planetary-mass objects and reveal correlations related to temperature and surface gravity. We show in Fig.~\ref{fig:color_magnitude} the absolute $H$ magnitude as function of the $H$~--~$M'$ color. This CMD was created with the \texttt{species} toolkit \citep[see also][]{stolker2020a} and includes a sample of \added{regular field dwarfs and young/low-gravity objects} \citep{dupuy2012,dupuy2013,liu2016}, directly imaged planets and brown dwarfs \citep{garcia2017,biller2010,biller2012,daemgen2017,chauvin2017,milli2017,ireland2011,bailey2013,bonnefoy2014,currie2012,currie2013,galicher2011,lacour2016,rajan2017,stolker2019,stolker2020b,stolker2020a}, and synthetic photometry calculated from \added{the AMES-Cond/Dusty} isochrones at \replaced{5~Myr}{3~Myr} and model spectra \citep{chabrier2000,allard2001,baraffe2003}, \added{that is, hot-start cooling tracks with a clear or cloudy atmosphere}. We note that narrowband SPHERE $H2$ magnitudes are shown for HIP~65426~b and HD~206893~B, so ignoring minor color effects. \added{The very red $H$~--~$M'$ colors of these two objects are probably caused by an enhanced dust content in their atmospheres \citep{cheetham2019,stolker2020a} although a scenario in which additional reddening is caused by circumplanetary material could also be considered given their age uncertainties.}

For \gq, we computed synthetic $H$-band photometry from the VLT/SINFONI spectrum that will be analyzed in Sect.~\ref{sec:modeling}. The MKO $H$-band photometry is $14.02 \pm 0.13$~mag so the $H$~--~$M'$ color is $2.04 \pm 0.15$~mag. The error is dominated by the uncertainty on the \emph{HST}/NICMOS F171M flux from \citet{marois2007} that was used to calibrate the $H$-band spectrum. A comparison with the spectral sequence of the high-gravity field dwarfs shows that the absolute $H$-band brightness of \gq is consistent with a mid M-type object, somewhat similar to PZ~Tel~B and HD~1160~B. When comparing with the AMES-Cond/Dusty isochrones, we estimated a photometric mass for \gq of $\sim$30~\MJ at an assumed age of 3~Myr.

The color of \gq is $\sim$1~mag redder than the synthetic predictions from the atmosphere models and $\sim$1.2~mag redder than the field dwarfs of similar spectral type. For a mid M-type object, the atmosphere is expected to be cloudless, which can indeed be seen from the convergence of the AMES-Cond and AMES-Dusty isochrones in Fig.~\ref{fig:color_magnitude} for masses $\gtrsim$20~\MJ. Given the previously detected hydrogen emission lines \citep{seifahrt2007,zhou2014,wu2017a}, the red $H$~--~$M'$ color \replaced{likely points to}{is likely caused by} thermal emission from dust grains in a protolunar disk, that is, a disk around a planetary- or substellar-mass companion in which satellites might be forming \citep[e.g.,][]{wu2017a,perez2019}.

The color of \gq is comparable to the young ($\lesssim$10~Myr), planetary-mass objects GSC~06214~B and ROXs~42~Bb. It has been suggested that GSC~06214~B hosts a dusty accretion disk, as inferred from the Pa$\beta$ \citep{bowler2011} and H$\alpha$ emission \citep{zhou2014}, and the red SED colors \citep{bailey2013}. Hydrogen emission lines have not been identified in the spectrum of ROXs~42~Bb \citep{bowler2014} and there seems no hint of excess emission from a disk although the SED is reddened by $A_V \approx 1.7$--$1.9$~mag \citep{currie2014a,bowler2014}. In contrast to \gq, the red colors of these early L-type objects may also point to enhanced cloud densities in their low-gravity atmospheres. Indeed, the AMES-Dusty predictions (i.e., with strongly mixed clouds) are consistent with GSC~06214~B and ROXs~42~Bb within 1--3$\sigma$.

\replaced{While some extinction is to be expected for \gq in case the companion is surrounded by an inclined disk, a reddening of the $H$~--~$M'$ color by $\sim$1~mag would require a rather large extinction of about $\sim$2~mag in the $H$ band when considering submicron-sized dust grains (see arrows in Fig.~\ref{fig:color_magnitude}). A large extinction by small dust grains is therefore not expected to be the origin of the red $H$~--~$M'$ color given the clear detection at optical wavelengths.}{Thermal emission \added{from a disk} may cause a red $H$~--~$M'$ color, but extinction alone could in principle have a similar effect. With the presence of an inclined disk, some extinction is expected to occur, in particular in the surface layer where presumably submicron-sized grains reside. In Sect.~\ref{sec:model_atmos}, we will estimate a visual extinction of $A_V = 2.3$ by modeling the optical and NIR spectra of \gq with an atmospheric model and assuming an ISM-like extinction, using the empirical relation from \citet{cardelli1989}. In that case, the reddening of $H$~--~$M'$ will be $\sim$0.4~mag (see arrow in Fig.~\ref{fig:color_magnitude}), which is a factor $\gtrsim$3 smaller than what has been measured. A much larger reddening is therefore required, but such a scenario seems unlikely given the constraints on the atmospheric parameters and the actual detection of \gq at optical wavelengths.}

\subsection{Optical spectral type and features}
\label{sec:muse_analysis}

\begin{figure*}
\centering
\includegraphics[width=\linewidth]{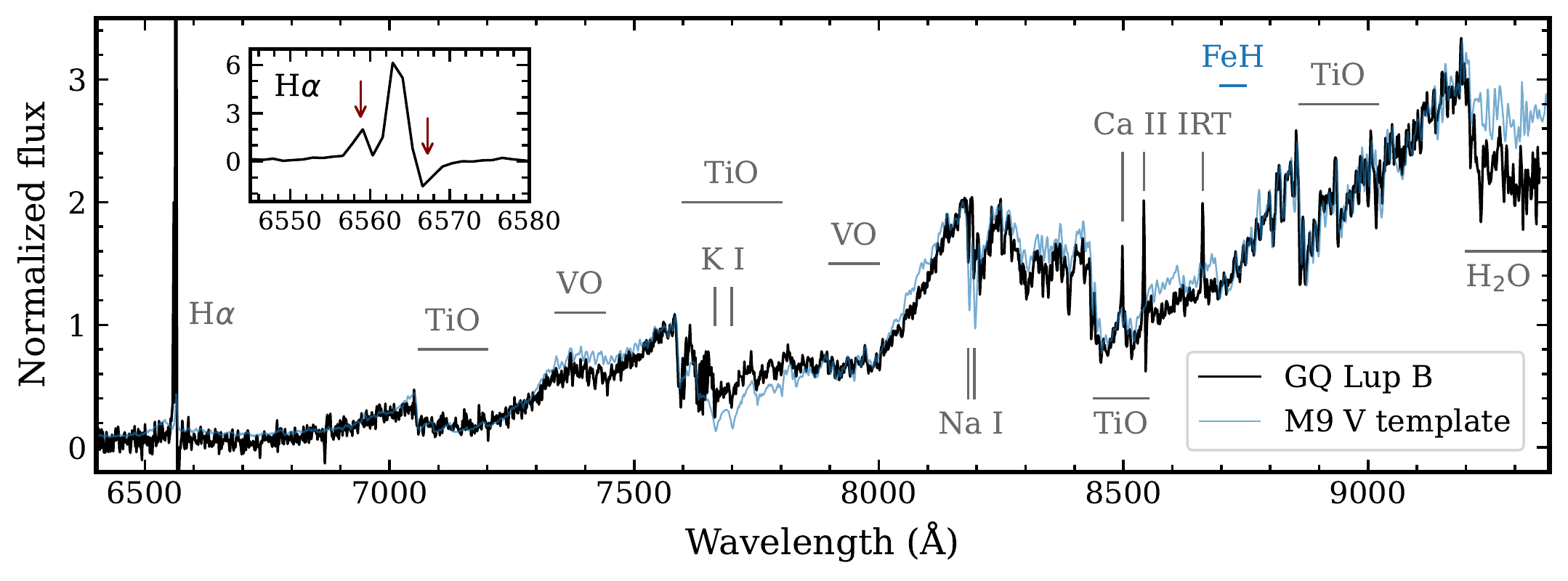}
\caption{Optical spectrum of \gq with MUSE. The extracted spectrum (\emph{black line}) is shown in comparison with a standard M9~V stellar template (\emph{blue line}). The opacity sources of the identified spectral features are labeled in \emph{gray} and \emph{blue} with the latter in case a species is only visible in the stellar template. \added{The \emph{inset} shows a zoom to the H$\alpha$ line with the \emph{arrows} pointing to the stellar residuals (see Sect.~\ref{sec:emission_lines} for details).}}
\label{fig:muse_spec}
\end{figure*}

Figure~\ref{fig:muse_spec} shows the extracted optical spectrum from MUSE, with the main spectral features labeled, compared to a main-sequence M9 spectral template from \citet{kesseli2017}. We estimated the optical spectral type of \gq by using \texttt{PyHammer} \citep{kesseli2017}, which relies on the comparison of spectral indices with templates, and found a best-fit spectral type between an M8 and M9. As a second method, we used spectral type relations that are specific for young stars and found a spectral type of M7.4 with the TiO~8465 index from \citet{herczeg2014}. This discrepancy of about one spectral subtype is expected and has been previously noted when comparing young pre-main-sequence objects to main-sequence stars \citep{herczeg2014}. With both methods we have not accounted for extinction (see Sect.~\ref{sec:model_atmos}) but the reddening mainly impacts the broadband SED slope while the spectral indices that were used for inferring the spectral type are not much affected.

The majority of the differences between the spectrum of \gq and the M9 template can be explained by the lower surface gravity of \gq. For example, we identify absorption from alkali metals, \ion{K}{1} and \ion{Na}{1}, but both doublets are narrower and weaker in \gq due to the decreased pressure broadening in low-gravity atmospheres \citep{allers2013}. The many TiO and VO bands dominate the spectrum and set the pseudo-continuum in both \gq and the main sequence template. However, in the regions around 7400 \AA\ and 8000 \AA, it seems that VO might be enhanced in \gq compared to the template, which is another indicator for a low surface gravity \citep{martin1996,mcgovern2004}. Finally, we do not detect any absorption from metal hydrides (FeH, CrH, or CaH) in \gq. This is in contrast to older directly imaged, planetary-mass objects where FeH is clearly detected (e.g., 2MASS~0103~(AB)b; \citealt{eriksson2020}), suggesting that FeH is not seen in the spectrum of \gq because its photosphere is located at low pressures. Metal hydrides have long been used as surface gravity indicators \citep[e.g.,][]{schiavon1997}, and the lack of their features is again complementary evidence for the youth and low-gravity nature of \gq.

\subsection{Hydrogen Emission Lines}
\label{sec:emission_lines}

\begin{deluxetable*}{lcccccc}
\tablecaption{Emission line measurements}
\label{table:emission_lines}
\tablehead{
\colhead{Line} & \colhead{$F_\mathrm{line}$} & \colhead{$L_\mathrm{line}$} & \colhead{EW} & \colhead{FWHM} & RV \\
 & \colhead{(W m$^{-2}$)} & \colhead{($\Lsun$)} & \colhead{(\AA)} & \colhead{(km s$^{-1}$)} & \colhead{(km s$^{-1}$)}
}
\startdata
H$\alpha$ & $(3.31 \pm 0.04) \times 10^{-18}$ & $(2.38 \pm 0.03) \times 10^{-6}$ & $\cdots$ & $108 \pm 1$ & $19 \pm 1$ \\
H$\beta$ & $< 2.7 \times 10^{-19}$ & $< 2.0 \times 10^{-7}$ & $\cdots$ & $\cdots$ & $\cdots$ \\
Pa$\beta$ & $(1.32 \pm 0.01) \times 10^{-18}$ & $(9.53 \pm 0.10) \times 10^{-7}$ & $-3.72 \pm 0.04$ & $237 \pm 3$ & $15 \pm 1$ \\
\ion{Ca}{2} $\lambda$8498 & $(3.35 \pm 0.56) \times 10^{-19}$ & $(2.49 \pm 0.42) \times 10^{-7}$ & $-1.78 \pm 0.30$ & $97 \pm 16$ & $-22 \pm 8$ \\
\ion{Ca}{2} $\lambda$8542 & $(4.40 \pm 0.16) \times 10^{-19}$ & $(3.26 \pm 0.12) \times 10^{-7}$ & $-2.29 \pm 0.08$ & $84 \pm 3$ & $-18 \pm 1$ \\
\ion{Ca}{2} $\lambda$8662 & $(4.06 \pm 0.36) \times 10^{-19}$ & $(3.01 \pm 0.27) \times 10^{-7}$ & $-1.67 \pm 0.15$ & $110 \pm 5$ & $2 \pm 2$
\enddata
\tablecomments{The measurements have not been corrected for extinction ($A_V = 2.3$~mag; see Sect.~\ref{sec:model_atmos}). The listed values for H$\beta$ are 1$\sigma$ upper limits.}
\end{deluxetable*}

In addition to the detection of molecular and atomic species, the optical spectrum reveals several emission lines which are also labeled in Fig.~\ref{fig:muse_spec}. Specifically, we detect a prominent H$\alpha$ line and the \ion{Ca}{2} infrared triplet (IRT), which are both spectral signatures for accretion \citep[e.g.,][]{white2003}. There was no detection of H$\beta$ but we derived a $1\sigma$ upper limit of $2.7 \times 10^{-19}$~W~m$^{-2}$ by injecting artificial sources at a range of position angles while avoiding bright diffraction residuals. Previously, Pa$\beta$ emission was identified by \citet{seifahrt2007} in the SINFONI spectrum but the line characteristics had not been analyzed. The NACO NB4.05 filter is used by the MIRACLES program to detect atmospheric emission at $\sim$4~$\mu$m but this narrowband filter is actually optimized for the detection of Br$\alpha$ emission, which could also be emitted by an accreting planet \citep{aoyama2020}. With the SED analysis later on in Sect.~\ref{sec:mir_excess}, we find no significant excess flux in NB4.05 that may point to hydrogen line emission so it is consistent with a non-detection of Br$\alpha$.

We inferred the emission line characteristics of the hydrogen (i.e., H$\alpha$ and Pa$\beta$) and \ion{Ca}{2} lines with the \texttt{species} toolkit \citep{stolker2020a}. First, the continuum flux was estimated by fitting a 3rd order polynomial to a smoothed version of the spectrum. This step was not required for the H$\alpha$ line since the continuum level is consistent with the noise floor. Second, we fitted a Gaussian function to the (continuum-subtracted) emission line by evaluating a Gaussian likelihood function and using uniform priors. The posterior distributions of the three parameters (amplitude, mean, and standard deviation) where sampled with the nested sampling Monte Carlo algorithm MLFriends \citep{buchner2016,buchner2019}, as implemented in \texttt{UltraNest} \citep{buchner2021}, and propagated into a line flux, equivalent width, FWHM, and radial velocity. The equivalent width could not be determined for H$\alpha$ since the continuum of \gq was not detected at those wavelengths.

The measured and derived properties of the emission lines are listed in Table~\ref{table:emission_lines}. With a resolution of $R \sim 2500$ ($=$120~km~s$^{-1}$) in Pa$\beta$ (see \citealt{seifahrt2007}), the line might be (marginally) resolved. The width of the H$\alpha$ line, on the other hand, is consistent with the instrument resolution of $R = 2516$ (i.e., 119~km~s$^{-1}$; see Fig.~B.3 in \citealt{eriksson2020}), considering that there were some stellar residuals due to local variations in the LSF, which may have led to slight inaccuracies with the spectral extraction of the H$\alpha$ line. These effects were difficult to correct \added{and caused under/over-subtraction of the stellar light on the blue/red side of the H$\alpha$ line (see inset in Fig.~\ref{fig:muse_spec})} so we made sure that these artificial features were excluded from the fit. There is some spread in the RV of the line centers compared to $v_\mathrm{sys} = -2.8 \pm 0.2$~km~s$^{-1}$ that was measured by \citet{schwarz2016}. This probably points to slight inaccuracies in the SINFONI and MUSE wavelength solutions although this is only at the $\sim$10\% level of their resolving power. Later on, we will interpret the hydrogen line emission in more detail but the \ion{Ca}{2} lines will not be further analyzed.

\subsection{Atmospheric and Disk Modeling}
\label{sec:modeling}

\begin{deluxetable*}{lcccccccccc}
\tablecaption{Spectral analysis of \gq with atmospheric models}
\label{table:model_fit}
\tablehead{
& \colhead{\Teff} & \colhead{\logg\tablenotemark{a}} & \colhead{\Rp} & \colhead{$a_J$\tablenotemark{b}} & \colhead{$a_H$\tablenotemark{b}} & \colhead{$A_\mathrm{V}$} & \colhead{\Tdisk} & \colhead{\Rdisk} & $\Delta G$\tablenotemark{c} \\
 & \colhead{(K)} & & \colhead{(\RJ)} & & & & \colhead{(K)} & \colhead{(\RJ)} & \\
\hline
Parameter range & 2000--3500 & 2.5--5.5 & 1--5 & 0.1--5 & 0.1--5 & 0--5 & 10--2000 & 5--1000 &
}
\startdata
\multicolumn{10}{l}{\emph{MUSE spectrum}} \\
BT-Settl & 2500 & 4.0 & 2.85 & -- & -- & -- & -- & -- & 0 \\
BT-Settl + ext. & 2700 & 4.0 & 4.13 & -- & -- & 2.7 & -- & -- & -156 \\
\hline
\multicolumn{10}{l}{\emph{SINFONI spectra}} \\
BT-Settl & 2600 & 4.0 & 3.55 & 1.84 & 1.19 & -- & -- & -- & 0 \\
BT-Settl + ext. & 2600 & 4.0 & 3.55 & 1.84 & 1.19 & 0.0 & -- & -- & 0 \\
\hline
\multicolumn{10}{l}{\emph{MUSE + SINFONI spectra}} \\
BT-Settl & 2350 & 4.0 & 3.93 & 1.68 & 1.06 & -- & -- & -- & 0 \\
BT-Settl + ext. & 2700 & 4.0 & 3.77 & 1.32 & 1.03 & 2.3 & -- & -- & -11587 \\
\hline
\multicolumn{10}{l}{\emph{MUSE + SINFONI spectra + 3--5~$\mu$m photometry + ALMA}} \\
BT-Settl + ext. + disk\tablenotemark{d} & 2700 & 4.0 & 3.77 & 1.32 & 1.03 & 2.3 & $461 \pm 2$ & $65 \pm 1$ & --
\enddata
\tablenotetext{a}{The surface gravity, \logg, was estimated from the gravity-sensitive alkali doublets in the MUSE and $J$-band spectra.}
\tablenotetext{b}{Scaling parameter to account for potential inaccuracies in the absolute flux calibration.}
\tablenotetext{c}{The relative goodness-of-fit is calculated for each combination of spectra separately.}
\tablenotetext{d}{The atmospheric, scaling, and extinction parameters were fixed to the best-fit values of fitting the MUSE and SINFONI spectra.}
\end{deluxetable*}

\subsubsection{Data and Model Preparation}
\label{sec:data_model_prep}

To quantify in more detail the atmospheric and disk characteristics of \gq, we have complemented our optical spectrum and 4--5~$\mu$m photometry with the VLT/SINFONI $JHK$-band spectra from \citet{seifahrt2007}, Subaru/CIAO $L'$ photometry from \citet{marois2007}, and the ALMA Band~6 and~7 upper limits from \citet{wu2017a} and \citet{macgregor2017}. We have also adopted optical and NIR photometry from \citet{marois2007} and \citet{wu2017a} but these have not been included in the analysis given their small constraining power with respect to the spectra. The SINFONI $J$-band spectrum has been flux calibrated with the $J$-band magnitude of 14.90 from \citet{mcelwain2007}, and the $H$- and $K$-band spectra with the \emph{HST}/NICMOS F171M and F215N magnitudes from \citet{marois2007}. For the error bars of the spectra, we have adopted the per pixel SNRs from \citet{seifahrt2007} (i.e., 100 in $J$ and 30 in $H$ and $K$) and sampled Gaussian errors. Additionally, we obtained a telluric spectrum with the \emph{SkyCalc} interface \citep{noll2012} and scaled the error bars with the reciprocal of the transmission to account for systematic errors by telluric lines.

The combined data were analyzed with the \texttt{species} toolkit by fitting a grid of BT-Settl spectra \citep{allard2012}, for which we adopted the CIFIST release that used the solar abundances from \citet{caffau2011}. BT-Settl is a 1D radiative-convective equilibrium model which accounts for non-equilibrium chemistry and cloud physics. Clouds are not expected to form though at the photospheric temperature of \gq, $\Teff \sim 2500$~K, and the carbon content will be mainly locked up in CO molecules. To model the ALMA fluxes, we extended the spectra into the \replaced{mm}{millimeter} regime by fitting the fluxes at $\lambda > 50$~$\mu$m with a powerlaw function in log-log space. With the grid of model spectra at hand, we carried out several comparisons with the optical and NIR data.

\subsubsection{Atmospheric Emission and Extinction}
\label{sec:model_atmos}

\begin{figure}
\centering
\includegraphics[width=\linewidth]{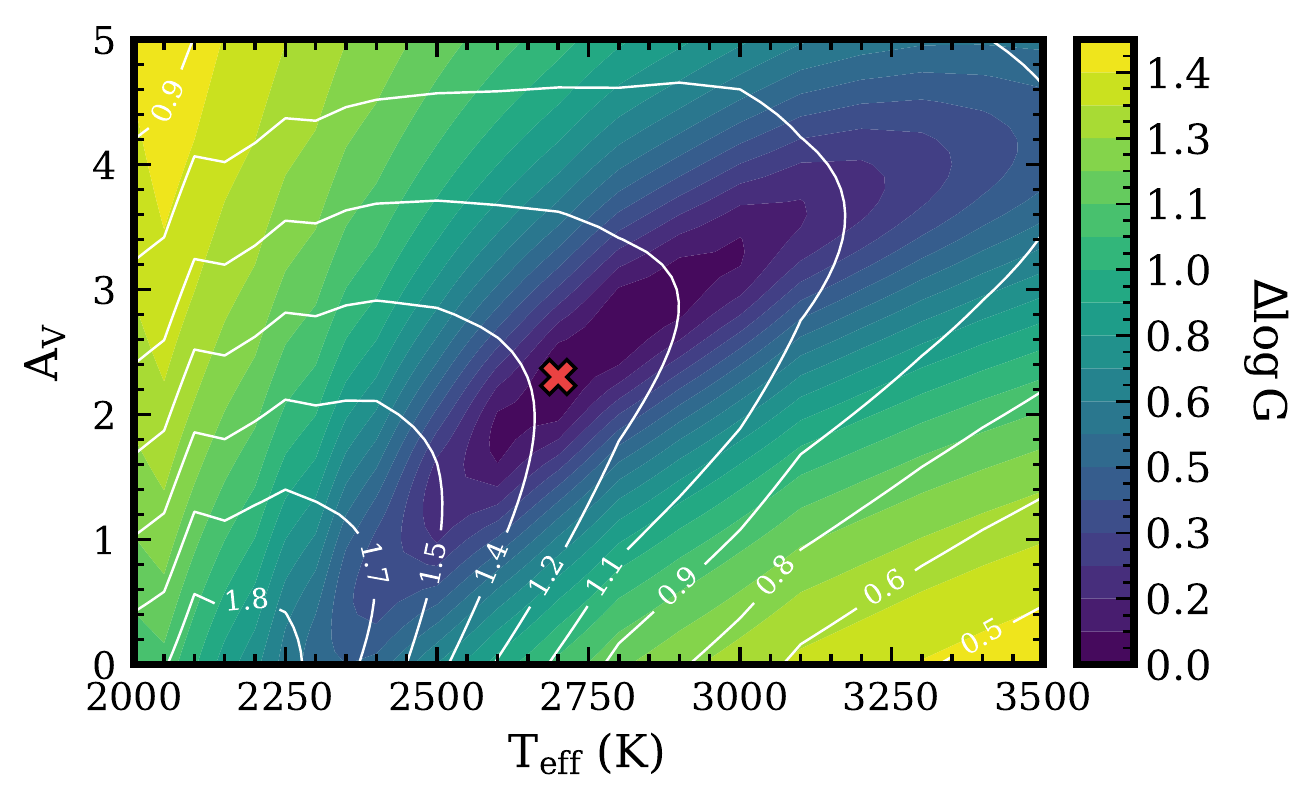}
\caption{Contours of the goodness-of-fit from the comparison of the MUSE and SINFONI spectra of \gq with the BT-Settl model grid. The \emph{color scale} shows the $G$ statistic relative to the best-fit parameters ($\Teff = 2700$~K, $\logg = 4.0$, and $A_V = 2.3$~mag), which are indicated with a \emph{red cross}. The \emph{white contours} show the best-fit scaling factor that is applied for recalibrating the $J$-band spectrum.}
\label{fig:model_grid}
\end{figure}

We first analyzed the MUSE and SINFONI spectra by considering atmospheric emission alone. This approach allowed us to constrain the main atmospheric parameters ($\Teff$, $\logg$, and $\Rp$) while assuming negligible emission from a disk up into the $K$ band. In order to account for potential inaccuracies with the absolute flux calibration of the spectra, we fixed the MUSE and $K$-band spectra (both were calibrated with \emph{HST} photometry) while fitting a scaling parameter, $a$, for the SINFONI $J$- and $H$-band spectra (the SINFONI spectra were obtained at separate epochs; see \citealt{seifahrt2007}). To account for interstellar and/or local (e.g., due to a disk) reddening, we adopted the empirical relation from \citet{cardelli1989} and include the visual extinction, $A_V$, as additional parameter while fixing the reddening to the standard value for the diffuse ISM, $R_V = 3.1$. Extinction in a disk is expected to occur mainly in the surface layer where grains are expected to be small. Hence, adopting an ISM relation might be a reasonable first choice for accounting for extinction even though the true grain properties may deviate. Therefore, we also tested a power-law parametrization (to account for the unknown dust properties) but obtained a similar result in which the wavelength-dependent extinction appeared comparable to the empirical ISM relation.

\begin{figure*}
\centering
\includegraphics[width=0.9\linewidth]{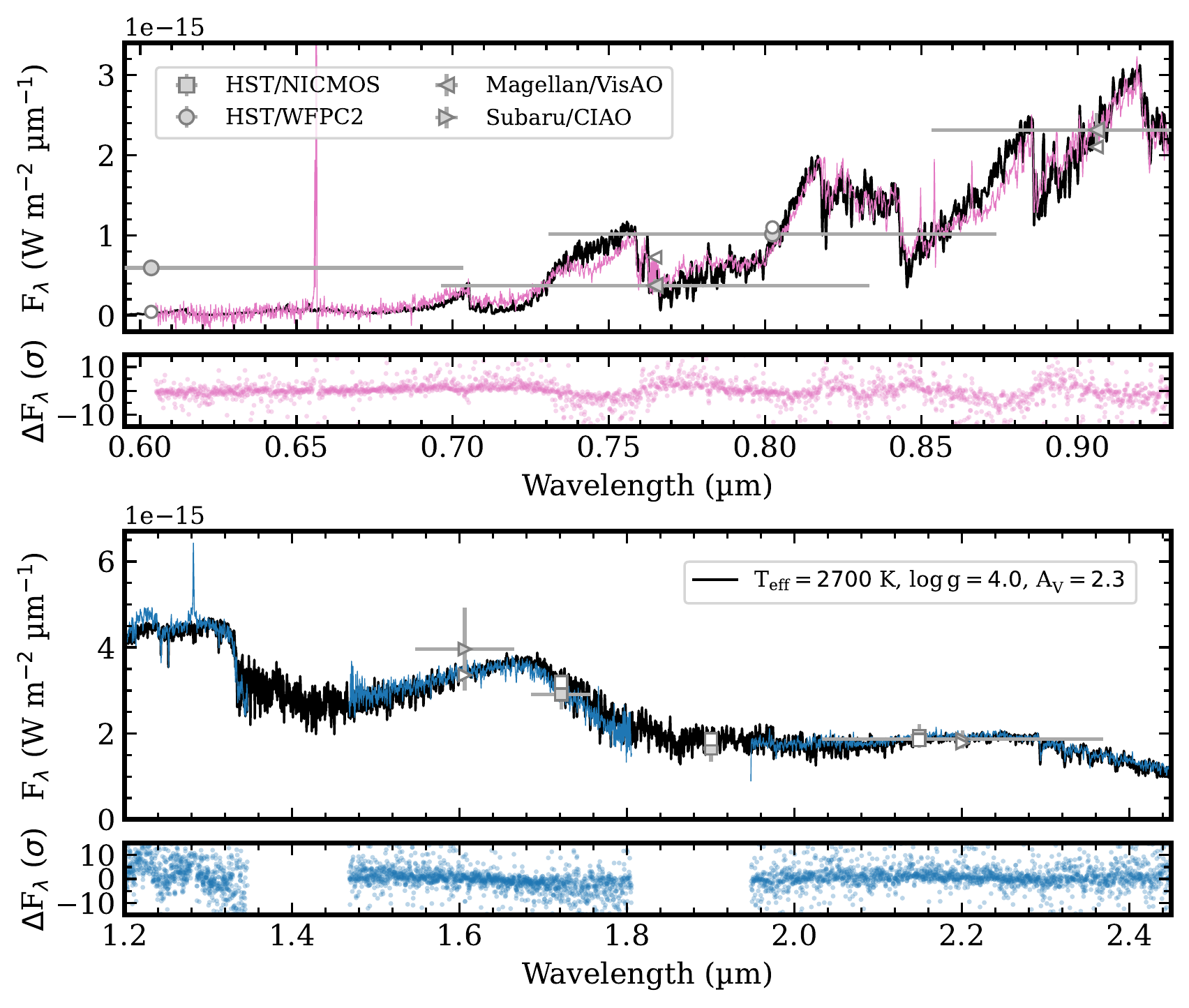}
\caption{Comparison of the best-fit model spectrum (\emph{black line}) with the MUSE (\emph{pink line; top panel}) and SINFONI spectra (\emph{blue line; bottom panel}) of \gq. The BT-Settl spectrum has been smoothed to $R = 3000$ but the residuals are shown at the actual resolution of each spectrum. For reference, we adopted the magnitudes from \citet{marois2007} and \citet{wu2017a} and converted these into fluxes (\emph{\replaced{square}{gray} markers}) with \texttt{species}. The horizontal error bars indicate the FWHM of the filter profiles and \emph{open \deleted{square} markers} are the respective synthetic photometry from the best-fit spectrum.}
\label{fig:muse_sinfoni}
\end{figure*}

The modeling approach that we first tested was a parameter retrieval which sampled spectra from the interpolated model grid and used Bayesian inference for estimating the uncertainties (i.e., similar as in \citealt{molliere2020} and \citealt{kammerer2021}). The residuals showed that the fit was limited by systematic errors in the models and/or data such that the uncertainties on the model parameters were highly underestimated. We tested different kernels for modeling the correlated noise with a Gaussian process but this only had a marginal effect on the width of the posterior distributions. At the level of precision and spectral resolution of the data, the fit was likely limited by inaccuracies caused by the subgrid sampling while spectral features are expected to change in a non-linear manner. A more sophisticated approach is required to account for such interpolation errors (e.g., \citealt{czekala2015}). In this work, we opt for a simplified approach by comparing the spectra from the discrete model grid and evaluating the $G$ statistic from \citet{cushing2008} (see their Eq.~1). It is mathematically a weighted $\chi^2$ statistic but it not expected to follow a $\chi^2$ distribution because of the systematic errors. As weights for the fluxes, we use the widths of the wavelength bins as suggested by \citet{cushing2008}.

We selected all available model spectra with $T_\mathrm{eff}$ in the range of 2000--3500~K with a 50~K sampling up to 2400~K and 100~K at larger temperatures. The grid spacing of the surface gravity, $\logg$, is 0.5~dex in the range of 2.5--5.5~dex. Additionally, we tested $A_V$ in the range of 0--5~mag in steps of 0.1~mag. Each model spectrum was first smoothed to the approximate resolving power of the instrument ($R = 3000$ for MUSE and $R = 2500, 4000, 4000$ for SINFONI $J$, $H$, and $K$) and then resampled to the wavelengths of the data. When comparing with the data, we fixed the distance and fitted the radius, $\Rp$, as flux scaling such that $G$ is minimized. Since optical and NIR wavelengths trace different parts of the atmosphere, we compare the model grid with the MUSE and SINFONI both separately and combined.

The broadband analysis of the SED did not constrain \logg since the impact of this parameter (at the considered \Teff) on the spectral fluxes is smaller than the typical systematic errors. The optical spectrum is however characterized by several low-gravity features (see Sect.~\ref{sec:muse_analysis}), which is further confirmed by the NIR spectra. We calculated \ion{K}{1}$_J = 1.05$, where \ion{K}{1}$_J$ is a gravity-sensitive spectral index that has been defined by \citet{allers2013}. The value is small compared to regular field dwarfs with a similar spectral type (see Fig.~22 in \citealt{allers2013}), confirming the low surface gravity. We determined, on visual inspection, that the optical and NIR alkali lines are best fitted with $\logg = 3.5$--4.0 when fixing $\Teff = 2700$~K. Specifically, the \ion{K}{1} doublet near 1.25~$\mu$m is best matched with $\logg = 4.0$ but the predicted \ion{K}{1} and \ion{Na}{1} lines at $\sim$0.77~$\mu$m and $\sim$0.82~$\mu$m, respectively, are a bit too strong and better matched with $\logg = 3.5$. \added{Similarly, with $\logg = 4.0$, the VO band at 0.74~$\mu$m is not sufficiently deep and it would require a lower surface gravity to match the model spectrum with the observed fluxes at those wavelengths.} Given the negligible impact on the overall fit, we set $\logg = 4.0$ in the remaining analysis.

The results from the spectral analysis are presented in Table~\ref{table:model_fit}. Several things can be noticed. Fitting the MUSE spectrum alone or combined with the SINFONI spectra yields a better fit when including $A_V$. This is seen from the last column in Table~\ref{table:model_fit} which lists the $G$ statistic relative to the case without extinction. The SINFONI spectra, on the other hand, do not constrain $A_V$. Instead, the $J$- and $H$-band scaling factors are relatively large so these parameters may mimic any reddening. Regarding \Teff, we obtained 2700~K when fitting the MUSE spectrum alone or combined with the SINFONI spectra. The NIR fluxes, however, were under/over-estimated when fitting the MUSE spectrum without/with extinction. Indeed, both \Rp and $A_V$ are different and expected to be more accurate when using spectra across a broad wavelength range.

When fitting the combined MUSE and SINFONI spectra, the $H$-band scaling is small and therefore consistent with the \emph{HST} photometry that was used for the calibration, although the residuals reveal a slope at the red end of the $H$ band (see Fig.~\ref{fig:muse_sinfoni}). This differential slope between model and data may point to an issue with the calibration of the pseudo-continuum (see also Fig.~5 in \citealt{lavigne2009} with a comparison of $JHK$ spectra from different instruments). The flux calibration of the $J$-band spectrum is expected to be the least accurate since we had adopted the photometric flux from a ground-based observation while the other spectra were calibrated with \emph{HST} photometry. When using all spectra and including $A_V$, the required flux scaling for the $J$-band spectrum is relatively large, $a_J = 1.3$ (see also contours in Fig.~\ref{fig:model_grid}). However, the $J$-band photometry was computed by \citet{mcelwain2007} while their spectra had been obtained at an airmass of 1.8 so a correction factor of 1.3 (i.e., $\sim$0.3~mag) seems reasonable.

All together, combining the MUSE and SINFONI spectra is expected to give the most accurate constraints. Therefore, the best-fit parameters from this work are $\Teff = 2700$~K, $\logg = 3.5$--$4.0$, $\Rp = 3.8~\RJ$, and $A_V = 2.3$~mag, which is shown in comparison with the spectra and some of the available photometric fluxes in Fig.~\ref{fig:muse_sinfoni}. We can also compare the derived atmospheric parameters with predictions from evolutionary models. At an age of 3~Myr, the AMES-Dusty isochrones \citep{chabrier2000} predict that a mass of $\Mp = 32~\MJ$ has $\Teff \approx 2700$~K, $\logg = 3.8$ (and therefore $\Rp = 3.7~\RJ$), and $M_H = 8$~mag. While hinging on the uncertain age estimate, this is in agreement with the inferred parameters from the spectra, as well as the absolute $H$-band brightness (see Fig.~\ref{fig:color_magnitude}). With the estimated $A_V = 2.3$, the extinction in the $H$ band is $\approx$0.4~mag so the object would be a bit brighter than the considered model prediction at 3~Myr \added{(see reddening vector in Fig.~\ref{fig:color_magnitude})}. This might be reasonable though given the margins of uncertainties and assumptions with the mass estimate.

While the best-fit model matches well with the overall morphology and spectral features, there are some systematic variations in the residuals. Apart from the discrepancy in the $H$ band, the optical absorption bands are somewhat muted compared to the model spectrum. Possibly this is caused by an inaccuracy with the calibration of the pseudo-continuum. Alternatively, the effect could be real if the absorption features are veiled by excess emission that originates from accretion hot spots \citep[e.g.][]{calvet1998}. It remains to be investigated if such continuum emission is to be expected at $\sim$0.7--0.9~$\mu$m. Differences between the MUSE spectrum and the MagAO photometry are likely attributed to the variability of GQ~Lup ($\Delta R \approx 0.4$~mag, $\Delta I \approx 0.3$~mag; \citealt{broeg2007}) since these fluxes were calibrated relative to the star (see \citealt{wu2017a} for details). In Fig.~\ref{fig:sed_fit}, we show the SED from optical to MIR wavelengths, together with the best-fit model spectrum. For clarity, here every 50th wavelength point is used. Interestingly, the best-fit model underpredicts the 3--5~$\mu$m fluxes by $1.3\sigma$, $4.9\sigma$, and $7.3\sigma$ in $L'$, NB4.05, and $M'$, respectively, when considering atmospheric emission alone (see dotted line in Fig.~\ref{fig:sed_fit}) which points to an additional luminosity component in the SED at those wavelengths.

\subsubsection{Thermal Emission from a Dusty Disk}
\label{sec:model_disk}

\begin{figure*}
\centering
\includegraphics[width=\linewidth]{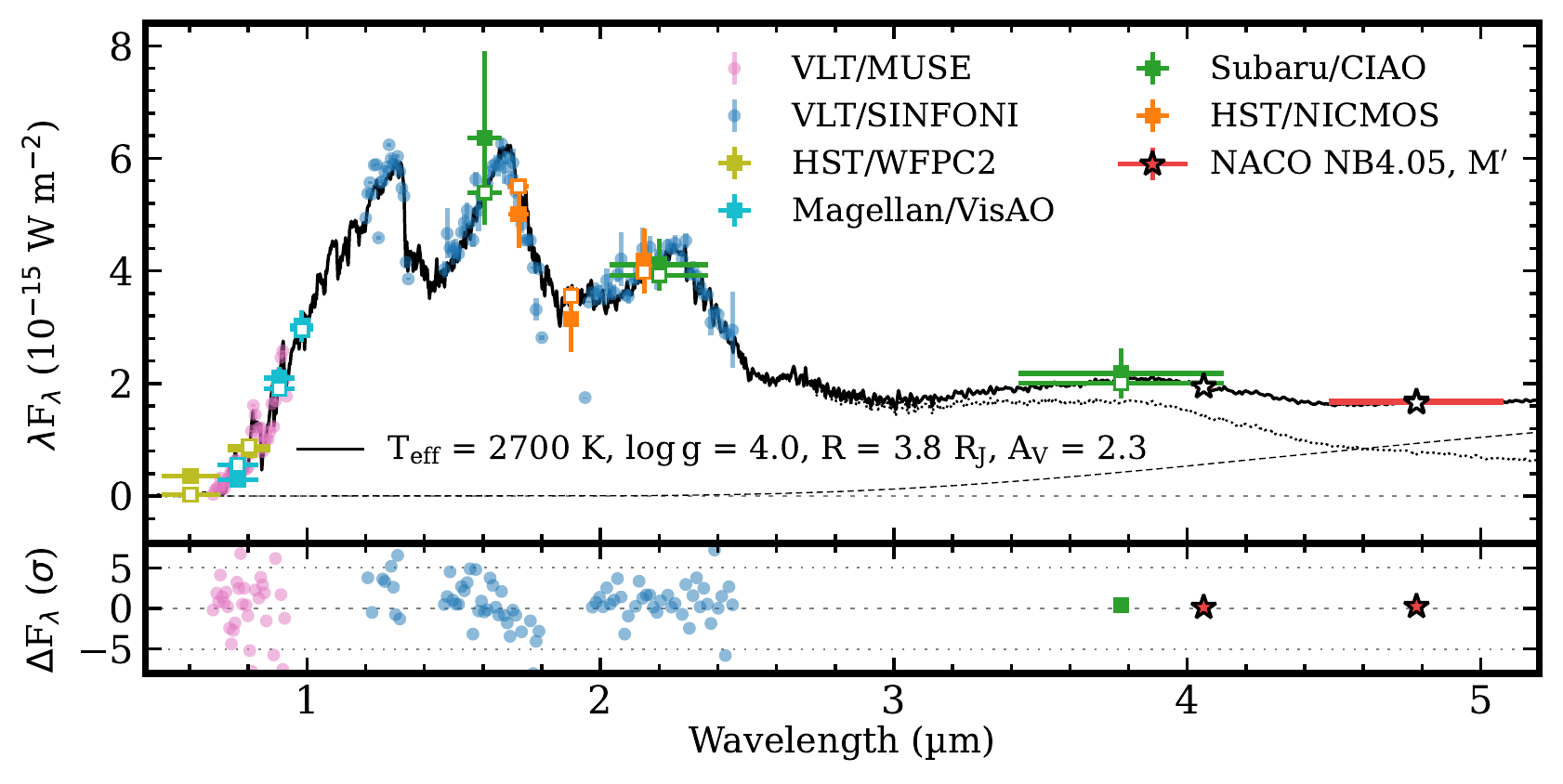}
\caption{Spectral energy distribution of \gq. The various spectral and photometric data points are shown with different colors as indicated in the legend. The NACO NB4.05 and $M'$ fluxes are highlighted with a \emph{red star}. The \emph{black solid line} is the best-fit model (smoothed to $R = 500$), including the additional blackbody component and the extinction applied. The \emph{black dotted} and \emph{black dashed} lines are the spectra of the atmospheric and blackbody component, respectively. The \emph{open markers} are the synthetic fluxes for all filters, with the same \added{shape and} edge color as the empirical fluxes. The \emph{lower panel} shows the residuals (i.e., data minus model) for the data points that were used in the fit.}
\label{fig:sed_fit}
\end{figure*}

The comparison with the atmospheric model spectra revealed excess emission at 3--5~$\mu$m. While there is some uncertainty in the absolute calibration of the $J$-band spectrum in particular, the \emph{HST} photometry by \citet{marois2007} is expected to be most accurate (ignoring possible variability of \gq) so we do not expect any spurious reddening in the $M$ band. Instead, the youth of \gq, the detected hydrogen emission lines, and the red NIR~--~$M'$ color all point to the presence of a protolunar disk.

To explore this hypothesis, we model the optical to MIR SED with a combination of atmospheric emission and a blackbody component that accounts for thermal emission from a dusty disk. To do so, we fix all parameters to the best-fit values from fitting the MUSE and SINFONI spectra and include, in addition to the spectra, the $L'$, NB4.05, and $M'$ fluxes, and the ALMA upper limits in the fit. We retrieve the effective \added{disk} temperature, $\Tdisk$, and radius, $\Rdisk$, with the Bayesian framework of the \texttt{species} toolkit for which we made again use of the nested sampling implementation of \texttt{UltraNest} \citep{buchner2021}. We used a Gaussian likelihood function for the model evaluation while mapping out the posterior space with 1000 live points and assuming uniform priors for both parameters (see Table~\ref{table:model_fit}).

The best-fit model spectrum, with the atmospheric and blackbody emission combined, is shown in Fig.~\ref{fig:sed_fit} and the posterior distributions of the disk parameters can be found in Fig.~\ref{fig:disk_posterior}. The modeled disk emission appears at wavelengths longer than $\sim$3~$\mu$m with indeed negligible flux in the $K$ band, as was to be expected from the residuals in Fig.~\ref{fig:muse_sinfoni}. The residuals of the $L'$, NB4.05, and $M'$ fluxes are $\leq 0.5\sigma$, indicating a good fit with a blackbody spectrum. For NB4.05, the transmission of the filter is optimized for the detection of Br$\alpha$ but there is no indication of such emission from \gq. Instead, the fit shows that the NB4.05 flux is consistent with continuum emission alone.

The retrieved parameters for a blackbody model are $\Tdisk = 461 \pm 2$~K and $\Rdisk = 65 \pm 1~\RJ$ (see bottom row in Table~\ref{table:model_fit}). From this, we derive a disk-to-companion luminosity ratio of $\approx$0.25, which implies that 25\% of the atmospheric emission may get reprocessed by the disk if we ignore additional processes (e.g., due to accretion) that may heat the disk internally. At \replaced{mm}{millimeter} wavelengths, \citet{wu2017a} and \cite{macgregor2017} reported an rms noise level of 39~and 50~$\mu$Jy~beam$^{-1}$ in Band~6 ($\lambda = 1.3$~mm) and~7 ($\lambda = 870$~$\mu$m), respectively. The synthetic ALMA Band~6 and~7 fluxes from the fit are $3.8 \times 10^{-24}$ and $1.8 \times 10^{-23}$~\flux, that is, \replaced{a factor 51 and 4 below}{a factor 18 and 11 below}, and therefore consistent with, the ALMA upper limits. Finally, resolving a disk radius of $65~\RJ$ would require a sub-mas resolution so the approximate radius that is probed by our observations is too compact to be resolved with ALMA.

\subsection{Orbital Architecture}
\label{sec:orbit}

\begin{figure*}
\includegraphics[width=\linewidth]{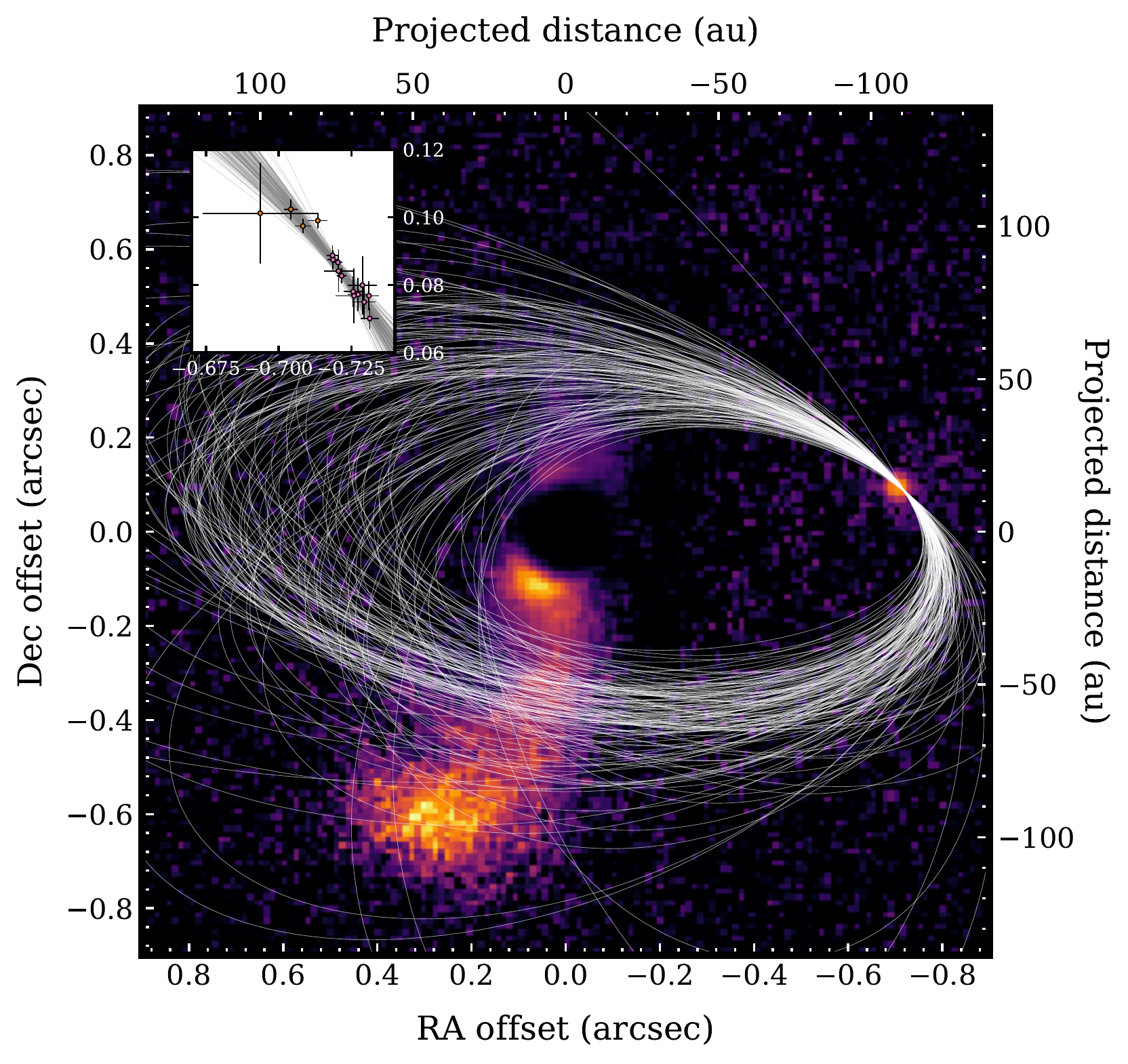}
\caption{Scattered light image of the circumstellar disk around GQ~Lup with sampled orbital configurations of \gq. The image shows the $r^2$-scaled, polarized flux on a linear color scale (see main text for details). The central part of the image ($\sim$0\ffarcs2 in diameter) was obscured by a coronagraph. The inset shows a zoom to the region of the astrometric data, obtained from 2004 to 2019. The \emph{orange markers} are the four astrometry points from this work. North is up and east is left.}
\label{fig:qphi_orbits}
\end{figure*}

The orbital architecture of \gq may provide clues about its formation and dynamical history. This is of particular interest because the polarimetric imaging observations by \citet{vanholstein2021} revealed an asymmetric, spiral-like structure in the CSD which points to a gravitational interaction by the companion. To estimate the orbital elements, we used \texttt{orbitize!}\footnote{\url{https://orbitize.readthedocs.io}} \citep{blunt2020} to fit the astrometry from \citet{neuhauser2008}, \citet{ginski2014}, and this work (see Table~\ref{table:astrometry}). The astrometric data are shown in the inset of Fig.~\ref{fig:qphi_orbits} with both the RA and Dec increasing over time. We also adopted the radial velocity (RV) of GQ~Lup, $v_\mathrm{sys} = -2.8 \pm 0.2$~km~s$^{-1}$ from \citet{schwarz2016} and their RV of \gq, \replaced{$-2.8 \pm 0.2$~km~s$^{-1}$}{$2.0 \pm 0.4$~km~s$^{-1}$}\explain{incorrect RV of the companion was shown here but the correct value was used in the orbit fit}, which breaks the 180~deg degeneracy in the longitude of the ascending node. The posterior distributions of the orbit parameters were sampled with a parallel-tempered MCMC algorithm \citep{foreman2013,vousden2016} while marginalizing over the uncertainty on the parallax ($6.49 \pm 0.03$~mas; \citealt{gaia2016,gaia2021}) and stellar mass ($1.03 \pm 0.05$~$\Msun$; \citealt{macgregor2017}).

For the parameter estimation, we used 20~temperatures, 500~walkers, and 50000~steps per walker. The first 40000~steps were removed as burn-in and we conservatively selected every 20th step of each walker to exclude correlations between steps. The remaining samples appeared converged on manual inspection, which was confirmed by calculating the integrated autocorrelation time of the chains. The posterior distribution of the orbit fit can be found in Fig.~\ref{fig:orbit_posterior} of the appendix. We obtained a constraint on the semi-major axis and eccentricity of $a = 117^{+24}_{-23}$~au and $e = 0.24^{+0.32}_{-0.17}$. The fit favors circular and low-eccentricity orbits although intermediate to high values are not excluded. The orientation of the orbital plane relative to the sky is given by the inclination and the longitude of the ascending node, for which we derived $i = 60^{+5}_{-9}$~deg and $\Omega = 265^{+19}_{-10}$~deg.

Figure~\ref{fig:qphi_orbits} shows the orbital constraints in comparison with the $Q_\phi$ image from \citet{vanholstein2021} that was obtained with VLT/SPHERE in the $H$ band (i.e., the same datasets that we used for the astrometry in Sect.~\ref{sec:archival_sphere}). The orbits were calculated by drawing 150~random samples from the posterior distribution. To enhance the visibility of the spiral arm (south in Fig.~\ref{fig:qphi_orbits}), we used \texttt{diskmap}\footnote{\url{https://diskmap.readthedocs.io}} \citep{stolker2016} to apply an $r^2$-scaling that accounts for both the orientation of the midplane, for which we adopted an inclination of $60.5 \pm 0.5$~deg and a position angle of $346 \pm 1$~deg from \citet{macgregor2017}, and a power-law approximation for the vertical extent of the disk surface. We note that the apparent, polarized flux of \gq is caused by the subtraction of the unresolved, polarized, stellar component while the actual companion is not polarized up to $\sim$0.2\% \citep[see][for more details]{vanholstein2021}.

With the detected spiral arm by \citet{vanholstein2021} and the kinematics from \citet{macgregor2017} we can uniquely determine the orientation of the disk by assuming that the spiral arm has a trailing morphology with respect to the rotation direction of the CSD. Since the red/blue-shifted velocities are located in north/south direction, this implies that near side of the disk is in the western direction. Knowing the inclination and longitude of the ascending node of the CSD, we constrained the mutual inclination between the orbit and the CSD to $84 \pm 9$~deg by propagating the posterior distributions from the orbital fit and accounting for the uncertainty on the disk orientation.

\section{Discussion}
\label{sec:discussion}

\subsection{Mid-infrared Excess from a Protolunar Disk}
\label{sec:mir_excess}

Infrared colors have been a powerful probe for the inference of CSDs around young stars for decades \citep[e.g.,][]{lada2000}. Surveys of young, substellar objects in nearby star-forming regions have also revealed that many show excess emission from a dusty disk \citep{liu2003,muench2001}. \citet{jayawardhana2003b} carried out a systematic study in the $L'$ band and reported that objects around the substellar boundary have inner disk lifetimes comparable to T~Tauri stars, indicating that isolated brown dwarfs form through similar mechanisms as solar-mass stars. Similar to brown dwarfs, also young, planetary-mass objects are known to host disks \citep[e.g.,][]{zapatero2007}. The thermal excess flux correlates with detected H$\alpha$ emission, pointing to a common origin of a dusty accretion disk \citep{liu2003}. For low-mass objects, the accretion rates are typically lower and the H$\alpha$ line profiles are more narrow compared to T~Tauri stars \citep{jayawardhana2003a}.

The detection of both MIR excess and a strong H$\alpha$ line in the SED of \gq seems in line with expectations from isolated, substellar objects. Yet, the formation mechanisms might be different and the spiral-arm interaction may affect the accretion and disk characteristics of \gq. Both the disk of GQ~Lup and \gq may have been truncated and/or perturbed by the dynamical interaction, which could be the origin of the somewhat large extinction, $A_V = 2.3$~mag. While the modeling in Sect.~\ref{sec:modeling} revealed a significant excess emission at NB4.05 and $M'$, the $1\sigma$ deviation from the expected atmospheric emission in $L'$ had in fact already been noted by \citet{marois2007}.

Thanks to the large separation and brightness of \gq, we were able to precisely measure the \wavel fluxes and estimate the blackbody parameters, while, in comparison, this was much more challenging for the closer-in planet PDS~70~b \citep{stolker2020b}. Assuming that the emission traces a single blackbody temperature (e.g., the strongly irradiated inner radius of the disk), we were able to constrain the disk parameters to $\Tdisk = 461 \pm 2$~K and $\Rdisk = 65 \pm 1~\RJ$ (see Sect.~\ref{sec:model_disk}). The three photometric fluxes appear well described by a blackbody spectrum although the true disk structure might be more complex (i.e., it will have a vertical and radial temperature gradient). In that regard, we note that the uncertainties on the disk parameters only reflect the statistical errors from the fit. The peak of the blackbody emission lies at $\approx$6--7~$\mu$m so it would make an appealing target for MIR spectroscopy with the \emph{James Webb Space Telescope} (\emph{JWST}) to investigate the disk and dust properties in more detail.

\subsection{Evidence for Satellite Formation?}
\label{sec:satellite_evidence}

The MIR excess that is inferred from the SED is expected to trace the thermal emission from (sub)micron-sized dust grains, which are heated by the radiation field of \gq and accretion processes within the disk. For a centrally-irradiated disk, the hottest and brightest part is the inner radius that is directly illuminated. The effective temperature, $\Tdisk = 461 \pm 2$~K, is therefore expected to trace the inner radius of the disk, which is much cooler than the effective temperature of the companion atmosphere, $\Teff \approx 2700$~K (see Sect.~\ref{sec:modeling}). This speculatively suggests that a large central cavity is present in the disk which has been created by some mechanism. The estimated disk radius, $\Rdisk = 65 \pm 1~\RJ$, may provide further evidence for a cavity since it is large compared to $\Rp \approx 3.8~\RJ$ from the atmosphere. While \Rdisk might be reasonable first-order estimate, a broader wavelength coverage and more detailed modeling are required to constrain the actual inner radius of the disk.

Several mechanisms can create a central cavity in the disk structure. The sublimation radius of the dust lies at $R_\mathrm{s} = \frac{1}{2}\sqrt{Q_\mathrm{R}}(\Teff/T_\mathrm{s})^2\Rp$, where $Q_R = Q_\mathrm{abs}(\Teff)/Q_\mathrm{abs}(T_\mathrm{s})$ is the ratio of the absorption efficiencies of the dust, $Q(T)$, for radiation at color temperature, $T$, of the incident and reemitted field \citep{tuthill2001}. When assuming blackbody grains, $Q_\mathrm{R} = 1$, with a sublimation temperature of $T_\mathrm{s} = 1500$~K, and adopting $\Teff$ and $\Rp$ of \gq, we obtain $R_\mathrm{s} \approx 6~\RJ$. For a relatively cool object such as \gq, $Q_\mathrm{R}$ is not expected to be larger than unity by a factor of a few (i.e., due to inefficient cooling), when considering different sizes of silicate grains \citep{monnier2002}. The central cavity that is inferred from the SED analysis can therefore not be explained by sublimation of silicate dust grains, which is also not surprising since $\Tdisk << T_\mathrm{s}$. Instead, $\Tdisk$ lies closer to the freeze-out location of H$_2$O, which would be located a bit further outward. Therefore, the H$_2$O ice line could potentially have provided an efficient growth mechanism for the assembly of satellites. \citet{heller2015} simulated the evolution of such ice lines in disks around super-Jovian gas planets. For their most massive case, 12~\MJ, the ice line would (at an early phase) be located at a comparable radius as inferred from the SED, but its location evolves strongly over time. While \gq is at least a factor 10 more massive than Jupiter, it is interesting to note that the four Galilean moons, which have orbits at $6$--$27~\RJ$, would all fit within the estimated disk radius, $\Rdisk \approx 65~\RJ$, so the cleared area could be sufficiently large for hosting a multi-satellite system.

Attempts with ALMA to detect thermal emission from pebble-sized grains at \replaced{mm}{millimeter} wavelengths yielded only non-detections, \replaced{that is}{specifically}, $3\sigma$ upper limits of 150~$\mu$Jy at 880~$\mu$m by \citet{macgregor2017} and 120~$\mu$Jy at 1.3~mm by \citet{wu2017b}. \citet{wu2017b} argued that the non-detections for \gq and other PMCs may imply that their disks could be very compact and optically thick in order to sustain a few \replaced{Myr}{megayears} of accretion. Alternatively, we propose a scenario in which the reservoir of mm-sized grains has actually been depleted by the formation of satellites and a central cavity may have opened within the inner disk structure. Large grains are expected to drift so they may have provided a continuous influx of solids for the assembly of satellites \citep{shibaike2017}. Without any pressures traps, pebbles would ultimately get accreted by \gq but the disk can sustain gas and small grains on longer timescales. The remaining gas gets accreted onto the companion with $\Mdot \approx 10^{-6.5}~\MJyr$ (see Sect.~\ref{sec:accretion_constraints}) while the disk may get replenished each time \gq crosses the plane of the CSD.

Alternatively, the central disk region may have been truncated by a magnetic field of \gq if such a magnetic field is sufficiently strong and the accretion rate is not too high \citep{lovelace2011}. In case of magnetospheric accretion, the H$\alpha$ emission may come from the hot spots on the atmosphere that are fed by accretion funnels from the disk. The magnetic field evolution of brown dwarfs and giant planets has been calculated by \citet{reiners2010}. For a $25~\MJ$ object, the predicated magnetic field strength is $B_\mathrm{p} = 2$~kG at an age of 3~Myr. Together with the estimated $\Rp = 3.8~\RJ$ and $\Mdot \approx 10^{-6.5}~\MJyr$, we calculated a truncation radius of $R_\mathrm{T} \approx 600~\RJ$ by using the relation from
\citet{lovelace2011} (see their Eq.~(2)). This truncation radius appears large compared to the $\Rdisk$ that we derived from the SED, but $B_\mathrm{p}$, $M_\mathrm{p}$, and $\Mdot$ are particularly uncertain. With the same parameters, but changing $B_\mathrm{p}$ to 40~G, we are in the situation where $R_\mathrm{T} \approx \Rdisk$. Therefore, a weak magnetic field may explain the inferred cavity size if at that radius material is able to couple with the field lines. Magnetospheric accretion is also expected to leave an imprint on the morphology of the hydrogen line profiles and would imply a small filling factor. There is no evidence for such effects in the current data (see Sect.~\ref{sec:accretion_constraints} for more details) but a confirmation at higher \added{spectral} resolution is required.

\subsection{System Architecture and Formation Pathway}
\label{sec:dynamical_evolution}

The orbital analysis in Sect.~\ref{sec:orbit} constrained the mutual inclination between the orbit of \gq and the CSD to $84 \pm 9$~deg. Similarly, \citet{wu2017a} already suggested that these two planes are possibly misaligned, as inferred from the orbital solutions by \citet{ginski2014} and \citet{schwarz2016}. The authors also pointed out that the CSD is misaligned with the stellar spin axis, which has an inclination of $27.5 \pm 5$~deg \citep{broeg2007}. This would not be unusual for a T~Tauri star with a strong magnetic field, but could in this case also have been induced by a torque from \gq. While the orbit has a near-polar configuration with respect to the CSD, the misalignment with the spin axis of the star would be different although still significantly misaligned.

A large mutual inclination may have two origins. If the companion has formed from the circumstellar disk then this points to a rather dynamical history in which, for example, a gravitational interaction with the CSD or a catastrophic event has placed \gq on a highly misaligned orbit. The second companion that was recently discovered at a projected distance of 2400~au may also have influenced the orbital configuration of \gq \citep{alcala2020}. A scattering event typically results in a high eccentricity \citep{scharf2009,nagasawa2011} which is not in agreement with the possibly low eccentricity that we inferred from the orbital fit (see Fig.~\ref{fig:orbit_posterior}). On the other hand, the large separation and moderate companion-to-star mass ratio may also imply a stellar-like formation pathway for \gq, which could also be the origin of the substantial mutual inclination \citep[e.g.,][]{jensen2014,bate2018}.

\added{Although the mutual inclination seems constrained from the orbital fit, some caution is required since modeling a small orbit coverage may lead to a bias in the inferred eccentricity and inclination \citep{ferrer2021}. From the posterior, we derived an orbital period of $\log{P/\mathrm{yr}} = 3.1 \pm 0.1$ so the astrometric points cover about $\sim$1\% of the full orbit. This may not be sufficient to robustly distinguish between an inclined orbit with a small eccentricity and a face-on orbit with a large eccentricity. On the other hand, the difference in RV between GQ~Lup~A and~B that was measured by \citet{schwarz2016} does exclude a face-on orbit. Higher-precision astrometric measurements, such as with VLTI/GRAVITY, may mitigate potential biases and enable more accurate constraints on the orbital parameters \citep{gravity2019}.}

In contrast to the companion's orbit, the inclination of its disk is not well constrained. \citet{vanholstein2021} obtained an upper limit of 0.2\% polarization on the scattered light flux from the disk of \gq. The authors suggested that the disk may have a moderate or low inclination, and/or has a low dust surface density (see their Fig.~14). If the disk of \gq is co-planar with its orbit, then the inclination would be $\approx$60~deg, as inferred with the orbital fit (see Fig.~\ref{fig:orbit_posterior}). That seems indeed a moderate inclination and may explain the polarization non-detection. On the other hand, the simulations by \citet{vanholstein2021} assumed a small inner radius of the disk (i.e., comparable to the dust sublimation radius) while a central cavity would decrease the fractional polarization because the scattered light flux is lower at a larger radius. Therefore, a large inner radius in a strongly inclined disk can not be excluded.

The extinction provides another constraint on the disk orientation. In Sect.~\ref{sec:model_atmos} we estimated $A_V \approx 2.3$~mag (i.e., an optical depth of $\tau \approx 2$) from the broadband SED, which is larger than what has been inferred for the star (e.g., $0.4 \pm 0.2$~mag; \citealt{seperuelo2008}). In case of \gq, the extinction may occur locally, for example in the disk surface, which would require that the disk is sufficiently inclined. Alternatively, the extinction might be caused by material that gets channeled by the spiral arm from the CSD to the companion. Any dust in this inflow of material could somewhat enshroud both the companion and its disk. A more detailed constraint on the disk inclination would be valuable to understand if gravitational torques from the star could have misaligned the disk \citep[e.g.,][]{batygin2012}---and therefore also the orbits of potential satellites---with respect to the rotation axis of \gq. If the companion has a strong magnetic field that is coupled to its disk then it might also be driving an inner disk warp. In that case, the disk could be misalignment with the orbital plane and may cause a variable extinction. High-precision photometric monitoring of \gq could potentially reveal such variable changes in the disk structure.

\subsection{Constraints on the Accretion Processes}
\label{sec:accretion_constraints}

\begin{figure*}
\centering
\includegraphics[width=\linewidth]{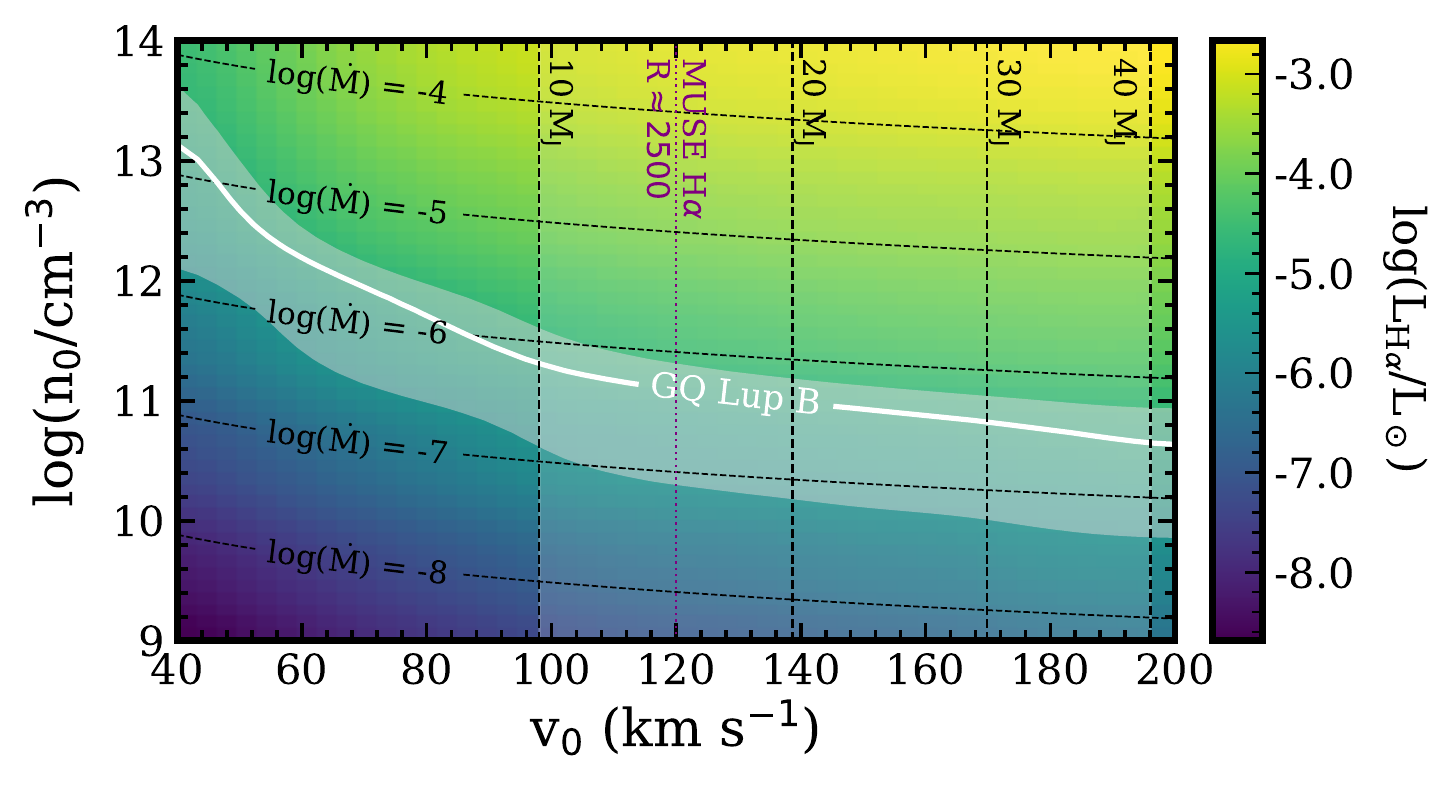}
\caption{Constraints on the mass accretion rate from the measured H$\alpha$ luminosity. The \emph{white contour} shows $L_\mathrm{H_\alpha}$ from \gq with an extinction correction of $A_V = 2.3$~mag and the surrounding \emph{shaded region} indicates $A_V$ in the range 0--3~mag. The \emph{horizontally dashed contours} are the accretion rates, $\log(\Mdot/(\MJyr))$, calculated from the model predictions and \gq parameters (see main text for further details), the \emph{vertically dashed lines} are the free-fall velocity for different masses of \gq, and the \emph{purple dotted line} is the resolving power of MUSE at H$\alpha$.}
\label{fig:accretion}
\end{figure*}

Hydrogen emission lines have been commonly used as tracers for accretion. Similar to stars, also brown dwarfs and giant planets radiate their accretion energy in the form of line and continuum emission \citep[e.g.,][]{zhou2014,zhu2015,aoyama2018}. Located in the Lupus~I cloud, \gq is a few million years old so previously detected hydrogen lines had been linked to the possible presence of disk around the companion. Several authors detected H$\alpha$ emission with ground- and space-based optical photometry \citep{marois2007,zhou2014,wu2017a}. The Pa$\beta$ line was found in the SINFONI spectrum by \citet{seifahrt2007} but interestingly the line was not detected in the $J$-band spectra by \citet{mcelwain2007} and \citet{lavigne2009}, even though the SNR was sufficiently high. This indicates that accretion rate \replaced{is}{might be} variable, for example due to density perturbations at the shock front, or possibly \gq even went through a quiescent state. Associated extinction effects that may occur in the vicinity of the atmosphere (e.g., by the disk and/or infalling material) may therefore be variable as well.

Line fluxes and ratios are a powerful diagnostic for the characterization of the accretion shock on a circumplanetary disk (CPD) or planet surface \citep{aoyama2020,hashimoto2020,uyama2021}. In Sect.~\ref{sec:emission_lines}, we measured the H$\alpha$ flux and estimated an upper limit for H$\beta$, yielding an extinction-corrected, $1\sigma$ upper limit on the line ratio of $F_{\mathrm{H}\beta}/F_{\mathrm{H}\alpha} < 0.17$. To quantify the accretion rate, we use the line predictions from \citet{aoyama2018} that were calculated with a detailed radiative hydrodynamical model of which the free parameters are the pre-shock velocity, $v_0$, and hydrogen number density, $n_0$. The line fluxes from the model are converted into $L_\mathrm{H_\alpha}$ by fixing the radius of the shock surface to $\Rp = 3.8~\RJ$ (see Sect.~\ref{sec:model_atmos}) and assuming a filling factor of $\ff = 1$. We can also map ($v_0$, $n_0$) to $\Mp$ (i.e., assuming that $v_0$ is equal to the free-fall velocity) and $\Mdot$ with Eqs.~7--9 from \citet{aoyama2020} such that the accretion rate can be extracted at the $L_\mathrm{H_\alpha}$ of \gq, which is shown Fig.~\ref{fig:accretion}. Later on, we will estimate a filling factor of $f_\mathrm{fill} > 0.6$ from the line ratio which would imply that \added{the actual} $\Mdot$ could be smaller by a factor $\sim$2.

To break the degeneracy between the shock density and velocity, we consider two constraints on the companion mass. First, in Sect.~\ref{sec:emission_lines}, we showed that the H$\alpha$ line is not resolved at the resolving power of MUSE (i.e., $\approx$120~km~s$^{-1}$ in H$\alpha$). If the gas reaches the planet surface with the free-fall velocity, this would point to a low mass of \gq ($\lesssim$15~\MJ; see Fig.~\ref{fig:accretion}), possibly making \gq a PMC, with an accretion rate of $\Mdot \approx 10^{-6.5}$--$10^{-6}~\MJyr$. However, as noted previously, there was possibly some over-subtraction of the stellar residuals at H$\alpha$ which could somewhat bias the line flux and width. Indeed, we measured a broader line for Pa$\beta$ ($\mathrm{FWHM} \approx 237$~km~$s^{-1}$) which could imply a higher mass object ($\Mp > 40~\MJ$). Figure~\ref{fig:accretion} shows that this mass lies at the edge of the grid and indicates an accretion rate that is a bit smaller, $\Mdot \approx 10^{-6.5}~\MJyr$. As a second approach, we adopted the mass estimate from Sect.~\ref{sec:model_atmos}, which we derived by comparing the atmospheric parameters and $H$-band magnitude with the AMES-Cond isochrone at 3~Myr, yielding $\Mp \approx 30~\MJ$. Therefore, the inferred mass from the Pa$\beta$ line and the $H$-band flux are roughly in agreement so we adopt $\Mdot \approx 10^{-6.5}~\MJyr$ as the estimate for the accretion rate.

With $A_V = 2.3$~mag, we corrected the H$\alpha$ luminosity to $L_\mathrm{H_\alpha} = 1.4 \times 10^{-5}~\Lsun$ (see Fig.~\ref{fig:color_halpha}), which is very similar to the (extinction-corrected) $L_\mathrm{H_\alpha} = 2.0 \times 10^{-5}~\Lsun$ from \citet{zhou2014} although the authors assumed $A_V = 1.5$~mag while their flux is a factor~6 smaller. The authors estimated a comparable accretion rate of $\Mdot = 10^{-6.3}~\MJyr$, which is interesting given the different modeling approaches that were followed. \citet{wu2017a} reported a smaller H$\alpha$ luminosity, $L_\mathrm{H_\alpha} = 10^{-5.9}$--$10^{-5.4}~\Lsun$, but using $A_V = 0.4$~mag so recalibrating to $A_V = 2.3$~mag would make the luminosity comparable to our finding. The authors estimated $\Mdot \approx 10^{-9}$--$10^{-8}~\MJyr$, using an empirical relation for T~Tauri stars, which is about two orders of magnitude smaller than the value derived in this work. Roughly speaking, the H$\alpha$ luminosity appears somewhat stationary, in contrast to the variable Pa$\beta$ line. Multi-epoch observations are required to investigate the accretion processes and variability in more detail, in which case it is recommended to target H$\alpha$ and Pa$\beta$ during the same night to mitigate potential biases with a multi-line interpretation.

A comparison of the upper limit on the line ratio, $F_\mathrm{H\beta}/F_\mathrm{H\alpha} < 0.17$, with the accretion model yields $v_0 \gtrsim 190$~km~s$^{-1}$ and $11.2 \lesssim \log{n_0/\mathrm{cm^{-3}}} \lesssim 11.8$. This agrees well with the constraint from the $L_\mathrm{H\alpha}$ that was shown in Fig.~\ref{fig:accretion} (i.e., the white contour at high $v_0$). The $n_0$ estimate from $L_\mathrm{H\alpha}$ is inversely proportional to $\ff$ because a smaller $\ff$ means a more concentrated accretion flow and a higher density at the shock. When $\ff < 0.6$, the $n_0$ derived from $L_\mathrm{H\alpha}$ is too dense to reproduce the low $F_\mathrm{H\beta}/F_\mathrm{H\alpha}$. Instead, the combined constraints from $L_\mathrm{H\alpha}$ and the line ratio point to a filling factor of $\ff > 0.6$. In contrast, $\ff \sim 0.01$ had been derived for PDS~70~b \citep{hashimoto2020} and 2MASS~0103~(AB)b \citep{eriksson2020} with the same analysis and also using MUSE data. Modeling of UV continuum for low-mass objects typically yields $\ff \lesssim 0.01$ \citep[e.g.,][]{herczeg2008}. Therefore, the large $\ff$ for \gq suggests a spherical-like accretion rather than concentrated accretion flow such as magnetospheric accretion. Possibly, the last crossing of \gq with the CSD caused an increased inflow of material which is being accreted with an approximate spherical geometry. This may also be the origin of the extinction if dust is entailed with the accretion flow.

\subsection{H$\alpha$--Color Characteristics of Directly Imaged Planets and Low-Mass Companions}
\label{sec:color_halpha}

\begin{figure*}
\centering
\includegraphics[width=\linewidth]{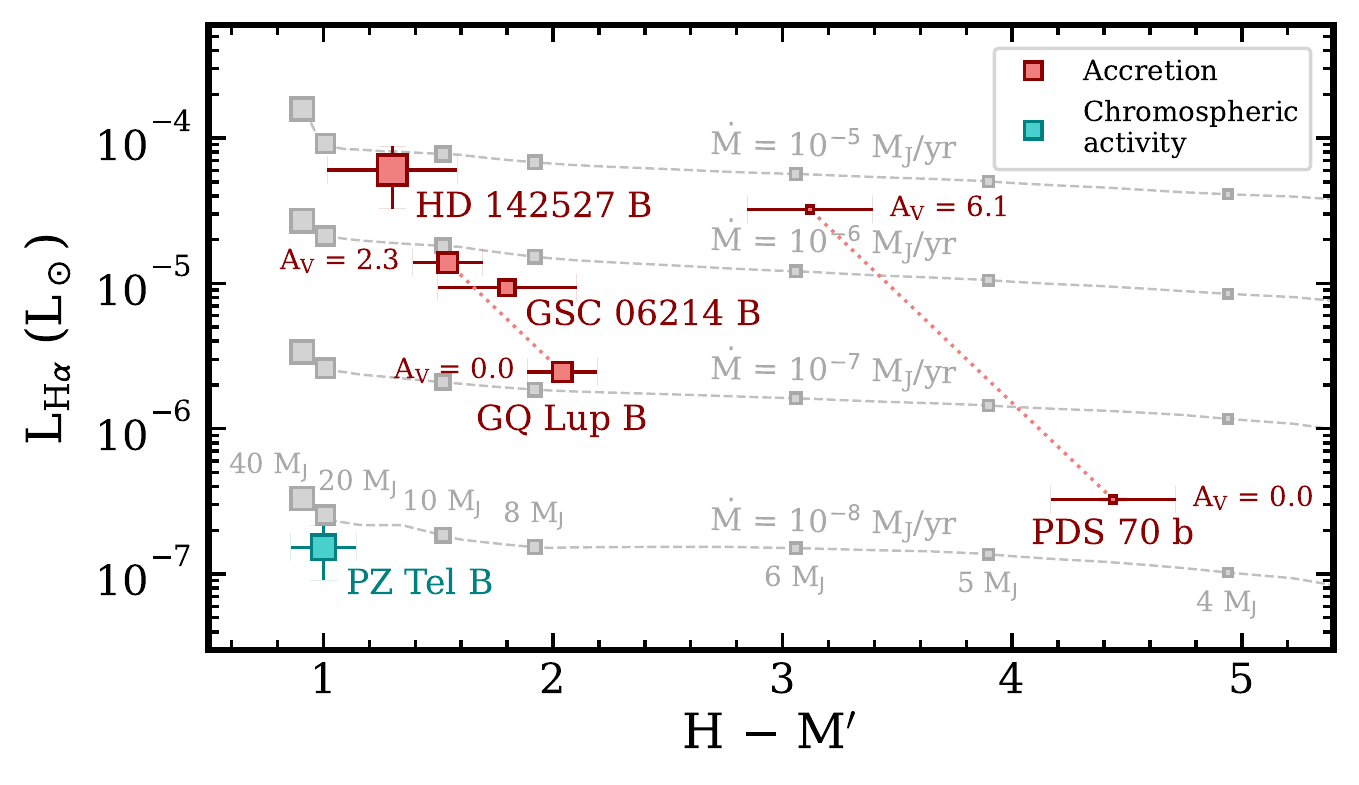}
\caption{Directly imaged, planetary and (sub)stellar companions with a measured H$\alpha$ luminosity and $H$~--~$M'$ color. The \emph{red markers} and \emph{teal markers} are objects for which the H$\alpha$ emission is expected to be associated with accretion and chromospheric activity, respectively. Model predictions (see main text for details) are shown with \emph{gray dashed lines} and marked at $4$, $5$, $6$, $8$, $10$, $20$, and $40~\MJ$ (\emph{from right to left}). The sizes of the \emph{squared markers} are all shown on the same logarithmic mass scale. For \gq and PDS~70~b, we show also the color--H$\alpha$ characteristics after dereddening with the $A_V = 2.3$~mag and $A_V = 6.1$~mag, respectively.}
\label{fig:color_halpha}
\end{figure*}

There is an increasing number of directly imaged planets and brown dwarfs with detected hydrogen emission lines. We can therefore start to empirically compare various types of objects to reveal possible trends in line luminosities, which may point to differences and commonalities in their formation mechanisms and accretion geometries. We therefore compiled a list of H$\alpha$ fluxes of directly imaged objects that also have $M'$ measurements. The $H$~--~$M'$ color is useful diagnostic for temperature/mass of a companion but is also affected the presence of a circumplanetary/substellar disk.

Figure~\ref{fig:color_halpha} shows $L_\mathrm{H\alpha}$ versus $H$~--~$M'$. Here, we have adopted the available magnitudes and line fluxes from \texttt{species} for PDS~70~b \citep{hashimoto2020,stolker2020b}, GSC~06214~B \citep{ireland2011,bailey2013,zhou2014}, PZ~Tel~B \citep{biller2010,musso2019,stolker2020a}, HD~142527~B \citep{biller2012,lacour2016,cugno2019}, and \gq (this work). Extinction effects are typically non-negligible during formation, especially at optical wavelengths, but can also be quite uncertain. We estimated $A_V \approx 2.3$~mag for \gq (see Sect.~\ref{sec:model_atmos}) but assuming that the absolute flux-calibration of the MUSE spectrum is accurate and that the object is not variable. Therefore, we show also the case of $A_V = 0.0$~mag for \gq in Fig.~\ref{fig:color_halpha}. Similarly, for PDS~70~b, we adopted $A_V = 6.1$~mag from \citet{wang2021}, which was shown to provide a good fit with \replaced{BT-Setll}{BT-Settl} spectra, but also show the $A_V = 0.0$~mag case for reference. For GSC~06214~B, we adopted $L_\mathrm{H\alpha} = 9 \times 10^{-6}~\Lsun$ from \citet{zhou2014}, which had been corrected for the assumed extinction of $A_V = 0.2$~mag.

With the extinction-corrected line luminosities, differences in origin of the H$\alpha$ emission become clear from Fig.~\ref{fig:color_halpha}. Late type field objects are known to be chromospherically active which produces somewhat low H$\alpha$ fluxes \citep[e.g.,][]{hawley1996,pineda2017}. The detected H$\alpha$ emission from PZ~Tel~B ($L_\mathrm{H_\alpha} \sim 10^{-7}$~$\Lsun$) is therefore expected to have a chromospheric origin instead of accretion, which seems also likely given the age of 24~Myr \citep{musso2019}. That interpretation is consistent with the lack of reddening in the $M'$ band (see \citealt{stolker2020a} and Fig.~\ref{fig:color_halpha}). The fraction of active M and L dwarfs peaks at the M/L transition \citep[e.g.,][]{west2004,schmidt2015}, so a small part of the H$\alpha$ flux of \gq may have a chromospheric origin though. The higher H$\alpha$ luminosities are instead expected to be associated with accretion. Indeed, these directly imaged objects in Fig.~\ref{fig:color_halpha} are all young (i.e., $\lesssim$10~Myr) with their host stars showing signs of accretion from a CSD. In case of \gq, the retrieved $\Teff = 2700$~K and $\Rp = 3.8~\RJ$ correspond to $\log(L_\mathrm{bol}/\Lsun) = -2.2$, such that $\log(L_\mathrm{H_\alpha}/L_\mathrm{bol}) = -2.7$. The latter is about an order of magnitude larger than typical luminosity ratios of late M-type field objects \citep{west2004}, while ignoring differences in age or mass. Indeed, \citet{barrado2003} derived a saturation level for chromospheric activity of $\log(L_\mathrm{H_\alpha}/L_\mathrm{bol}) = -3.3$ at early M-type T~Tauri stars. The detection of the \ion{Ca}{2} $\lambda$8662 line (see Sect.~\ref{sec:emission_lines}) further strengthens the interpretation, since this line is almost exclusively found in accreting objects while $\lambda$8498 and $\lambda$8542 may also have a chromospheric origin \citep{muzerolle2003,mohanty2005}.

In addition to the empirical data, Fig.~\ref{fig:color_halpha} shows model predictions that are derived by combining the AMES-Dusty isochrones and spectra \citep{chabrier2000,allard2001} with the hydrogen line predictions from the accretion model by \citet{aoyama2018}. Specifically, we extracted an isochrone at 5~Myr and interpolated $\Teff$, $\Mp$, and $\Rp$. From this, we computed synthetic photometry from the model spectra and, together with considered values for $\Mdot$, we calculated the pre-shock velocity and density. The latter were then used to interpolate the grid of H$\alpha$ fluxes that were also used in Sect.~\ref{sec:accretion_constraints}. The model predictions in Fig.~\ref{fig:color_halpha} show that lower mass objects are redder due to their lower temperature and emit less H$\alpha$ (assuming $\ff = 1$) since their radii and free-fall velocities are smaller. The accretion rate does in particular impact the H$\alpha$ flux with approximately a linear scaling between $L_\mathrm{H\alpha}$ and $\Mdot$ (see \citealt{aoyama2021} for a detailed analysis of the $\Mdot$--$L_\mathrm{H\alpha}$ relation).

The empirical data seems to follow the expectations from the numerical predictions. Lower mass objects (see symbol sizes in Fig.~\ref{fig:color_halpha}) are typically redder in $H$~--~$M'$ although dusty material in the vicinity could also redden a more massive object. In that regard, the color of the M-dwarf companion HD~142527~B is expected to be reddened by a circumsecondary disk \citep{lacour2016,christiaens2018}, similar to the detected excess flux from \gq. For PDS~70~b, it remains uncertain if the protoplanet is accreting from a CPD, in contrast to PDS~70~c which has been detected with ALMA \citep{benisty2021}. A slight, but not significant, excess flux was detected in $M'$ by \citet{stolker2020b} from which an upper limit of $\sim$256~K and $\sim$245~$\RJ$ was derived for potential emission from a CPD \citep[see also][]{christiaens2019}. The SED of PDS~70~b appears to be reddened as well and shows muted absorption features. Extinction effects are therefore expected to be non-negligible although the exact value remains under debate \citep{wang2021,cugno2021}. In general, it is also not clear if the extinction inferred from an SED is affecting the accretion luminosity in a similar way. A better understanding of the extinction properties will be key for deriving accurate line luminosities. Indeed, Fig.~\ref{fig:color_halpha} shows that a correction for extinction makes a large impact. The characteristics of \gq become similar to GSC~06214~B, and PDS~70~b is consistent with an even higher accretion rate. In fact, the extinction-corrected line luminosities from all considered, accreting, directly imaged objects are comparable within an order of magnitude even though they span over two orders of magnitude in mass.

Compared to PDS~70~b, \gq orbits further away from its primary star and is strongly misaligned with respect to the CSD (see Sect.~\ref{sec:orbit}). Hydrodynamical simulations have shown that wide-orbit, forming planets might be the origin of spiral structures that are seen in CSDs \citep[see e.g.,][]{dong2016}. Since \gq appears to drive a spiral arm, it remains to be further investigated whether the companion formed from the CSD or as a binary system by fragmentation from the same molecular cloud. Bottom-up formation timescale might be too long at $\sim$120~au but \citet{stamatellos2009} showed that large-separation brown dwarfs can form through disk fragmentation at an early phase. As already mentioned, some circumstellar gas may get channeled towards \gq and its disk, in particular when the companion crosses the CSD. ALMA observations by \citet{macgregor2017} resolved CO emission at the projected distance of \gq with a resolution of $\sim$0\ffarcs3. At higher resolution, ALMA may reveal a possible CO counterpart of the spiral arm, as well as kinematical perturbations in the vicinity of \gq, providing further insight into the most recent crossing with the CSD.

\section{Conclusions}
\label{sec:conclusions}

We have reported on the optical to MIR characterization of the directly imaged, substellar companion \gq. The optical spectrum is best matched with an M9 spectral type and contains several low-gravity indicators. We detected strong H$\alpha$ emission and derived an upper limit for H$\beta$. Analysis of the full SED, including the $JHK$-band spectra from \citet{seifahrt2007}, yielded $\Teff \approx 2700$~K, $\logg = 3.5$--$4.0$, $\Rp \approx 3.8~\RJ$, and a higher extinction, $A_V \approx 2.3$~mag, than previously estimated. The $H$-band flux and atmospheric parameters are consistent with $\Mp \approx 30~\MJ$ at an age of 3~Myr.

The $H$--~$M'$ color is $\approx$1~mag redder than mid/late field dwarfs due to an overluminosity at MIR wavelengths. A blackbody model provides a good fit to the 3--5~$\mu$m fluxes from which we derived $\Tdisk = 461 \pm 2$~K and $\Rdisk = 65 \pm 1~\RJ$, and is consistent with the ALMA upper limits. We used 15~yr of astrometric measurements to constrain the mutual inclination between the orbital plane of \gq and the CSD to $84 \pm 9$~deg and we tentatively showed that a low-eccentricity orbit is favored.

Our main conclusions are the following:

\begin{enumerate}

    \item[(i)] The MIR excess in the SED of \gq is caused by thermal emission from small dust grains in a protolunar disk.

    \item[(ii)] The H$\alpha$ line emission in combination with the MIR excess indicates that \gq is actively accreting from its disk. In addition, material may get delivered directly from the CSD, which could explain the $\ff \approx 1$ (i.e., a spherical accretion geometry) that we derived from the H$\beta$/H$\alpha$ upper limit.

    \item[(iii)] Analysis of the optical and NIR spectra suggests that the SED is reddened by $A_V \approx 2.3$~mag. The extinction is expected to occur in the vicinity of \gq, for example in a moderately inclined disk or by infalling material from the CSD which somewhat enshrouds \gq and its disk.

    \item[(iv)] The disk of \gq is cool compared to its atmosphere, as well as dust sublimation temperatures. We speculate that the disk is in a transitional stage in which the inner regions have been cleared by the formation of satellites while the remaining gas is being accreted onto the atmosphere with $\Mdot \approx 10^{-6.5}~\MJyr$. A depletion of the pebble reservoir would explain the non-detections with ALMA.

    \item[(v)] The large mutual inclination between the orbit and CSD could imply a tumultuous dynamical history (e.g., driven by the second, large separation companion in the system) which shares similarities with strong spin-orbit misalignments of hot-Jupiters. Alternatively, \gq may not have formed from the CSD but collapsed directly from the same parent cloud as GQ~Lup.

    \item[(vi)] Multi-line measurements during a single observing night are required to mitigate potential biases due to variability in the accretion \replaced{rate}{processes} and to take further advantage of the line-ratio diagnostics.

\end{enumerate}

We have presented the first detailed characterization of a disk around a directly imaged, low-mass companion, which highlights the valuable window for ground-based observations at \wavel. Measurements of a larger sample could reveal trends as function of companion mass and formation environment. \replaced{With ALMA, we predict that a factor 4 increase in sensitivity in Band~7 would be sufficient for detecting the disk of \gq at mm wavelengths.}{At millimeter wavelengths, we predict that a factor $\gtrsim$10 increase in sensitivity is required to detect the disk component of \gq. This will be challenging with the current capabilities of ALMA but might be in reach with the Next Generation Very Large Array \citep{rab2019,wu2020}.} Finally, the nearing launch of \emph{JWST} provides the exciting opportunity to reveal the spectral appearance of the protolunar disk at medium resolution and will guide detailed modeling efforts of its structure and dust properties.

\begin{acknowledgments}

We want to thank Andreas Seifahrt and Michael McElwain for sharing their near-infrared spectra of \gq. T.S.\ acknowledges the support from the Netherlands Organisation for Scientific Research (NWO) through grant VI.Veni.202.230. This work was performed using the ALICE compute resources provided by Leiden University. S.Y.H.\ acknowledges the support provided by NASA through the NASA Hubble Fellowship grant \#HST-HF2-51436.001-A awarded by the Space Telescope Science Institute, which is operated by the Association of Universities for Research in Astronomy, Incorporated, under NASA contract NAS5-26555. G.-D.M.\ acknowledges the support of the DFG priority program SPP 1992 ``Exploring the Diversity of Extrasolar Planets'' (KU~2849/7-1 and MA~9185/1-1), and from the Swiss National Science Foundation under grant BSSGI0$\_$155816 ``PlanetsInTime''. S.P.Q. and G.C.\ thanks the Swiss National Science Foundation for financial support under grant number 200021\_169131. Parts of this work have been carried out within the framework of the NCCR PlanetS supported by the Swiss National Science Foundation. J.B.\ acknowledges support by Funda\c{c}\~{a}o para a Ci\^{e}ncia e a Tecnologia (FCT) through research grants UIDB/04434/2020 and UIDP/04434/2020 and work contract 2020.03379.CEECIND.

\end{acknowledgments}

\facilities{VLT:Antu(NACO), VLT:Yepun(MUSE), VLT:Melipal(SPHERE)}

\software{\texttt{species} \citep{stolker2020a}, \texttt{orbitize!} \citep{blunt2020}, \texttt{PynPoint} \citep{stolker2019}, \texttt{IRDAP} \citep{vanholstein2020}, \texttt{diskmap} \citep{stolker2016}, \texttt{PyHammer} \citep{kesseli2017}}

\appendix

\section{Posterior Distributions}

\begin{figure*}
\centering
\includegraphics[width=\linewidth]{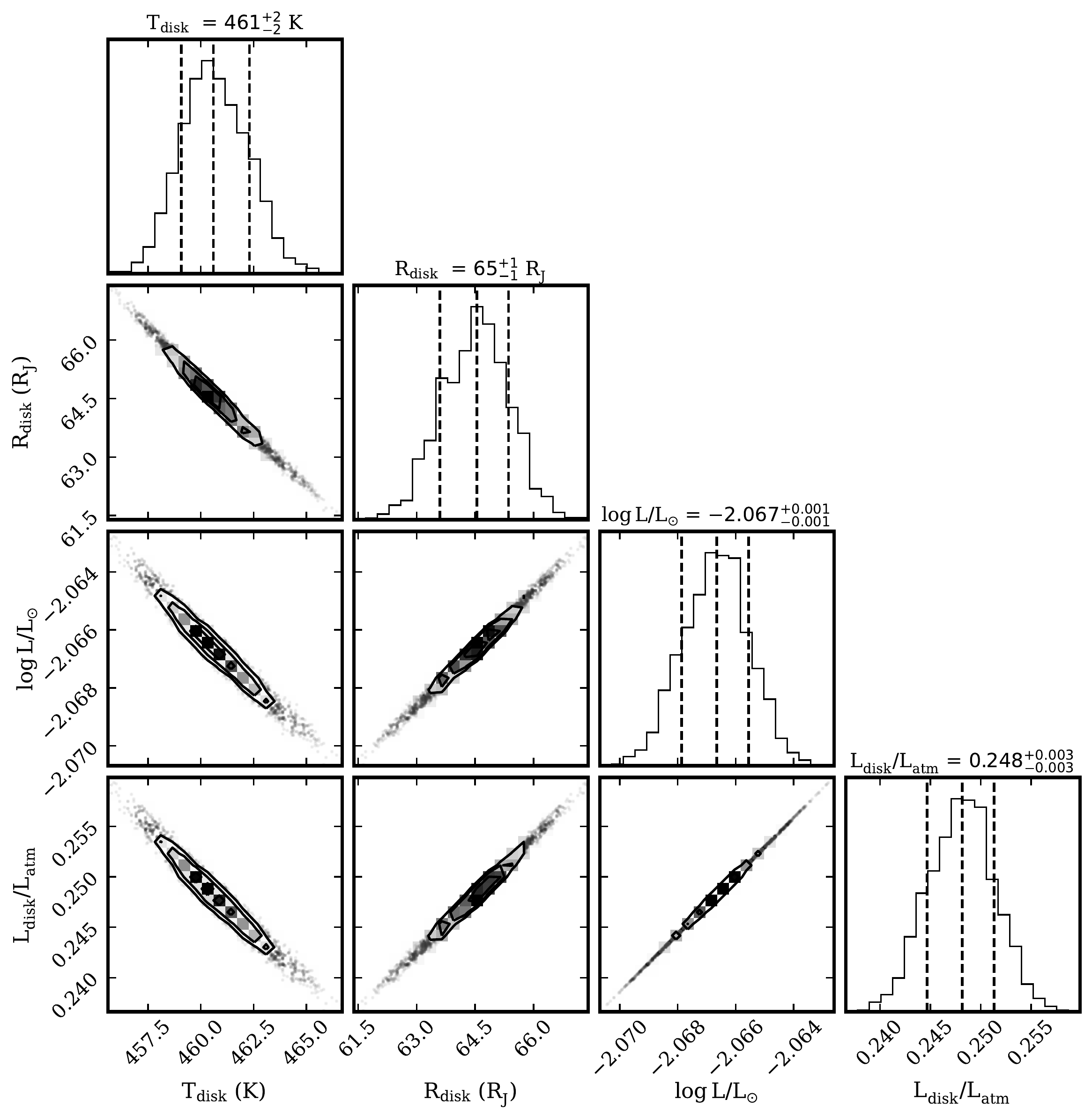}
\caption{Posterior distributions from fitting the \added{MUSE and SINFONI spectra, the photometric fluxes at 3--5~$\mu$m, and the ALMA upper limits} with a blackbody model while fixing the atmospheric parameters. The 1D marginalized distributions are shown in the \emph{main-diagonal panels} and the 2D projections of the posterior samples are shown in the \emph{off-diagonal panels}. The plot was created with \texttt{corner.py} \citep{foreman2016}. See Sect.~\ref{sec:modeling} for further details on the SED analysis.}
\label{fig:disk_posterior}
\end{figure*}

\begin{figure*}
\centering
\includegraphics[width=\linewidth]{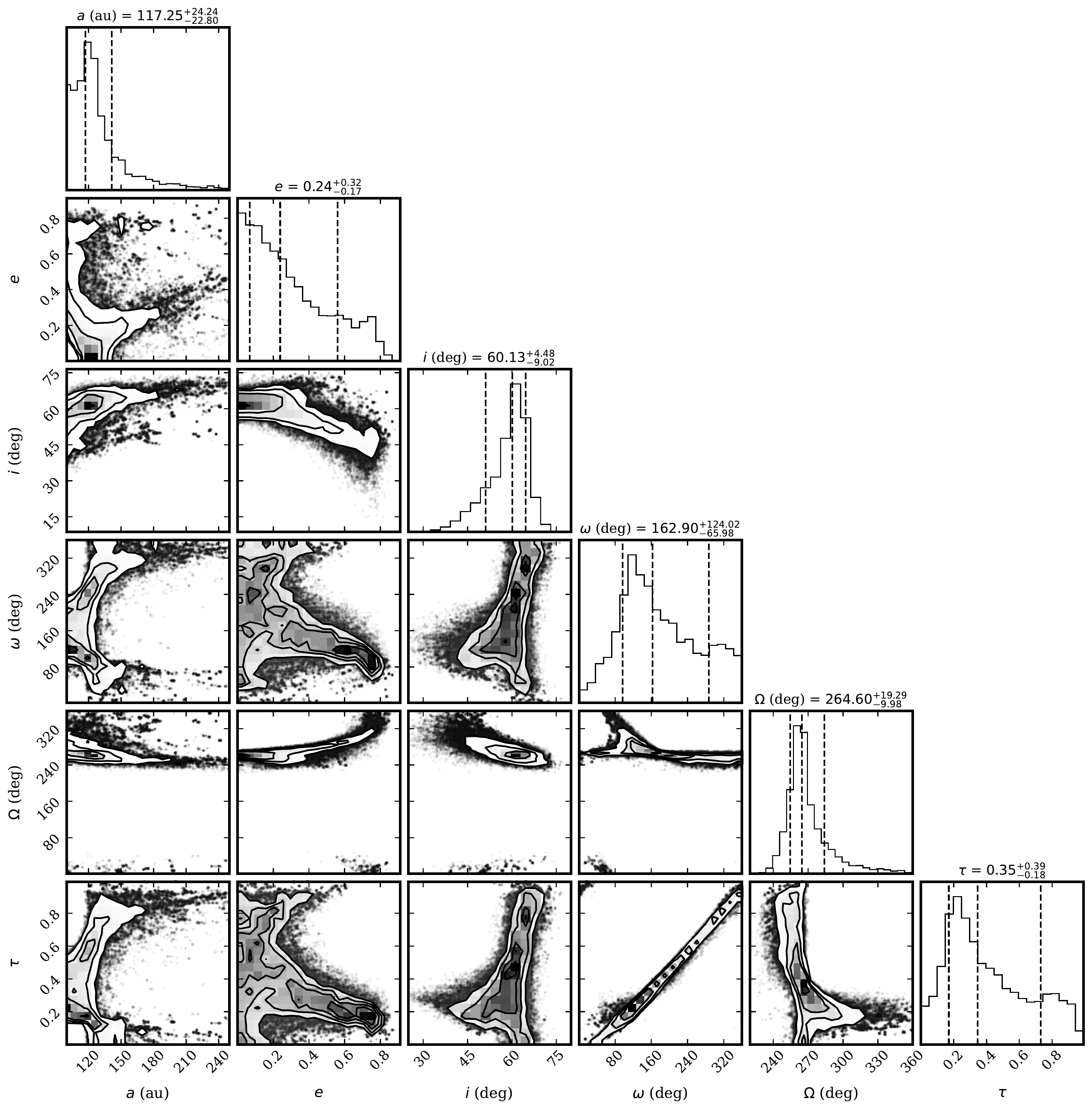}
\caption{Posterior distributions of the orbital elements of \gq. The 1D marginalized distributions are shown in the \emph{main-diagonal panels} and the 2D projections of the posterior samples are shown in the \emph{off-diagonal panels}. The plot was created with \texttt{corner.py} \citep{foreman2016}. See Sect.~\ref{sec:orbit} for further details on the orbit analysis.}
\label{fig:orbit_posterior}
\end{figure*}

\bibliography{references}
\bibliographystyle{aasjournal}

\end{document}